\documentclass[preprint]{aastex}

\def\aj{AJ}                   
\def\apj{ApJ}                 
\def\apjl{ApJ}                
\def\apjs{ApJS}               
\def\apss{Ap\&SS}             
\def\aap{A\&A}                
\def\mnras{MNRAS}             
\def\pasj{PASJ}               
\def\nat{Nature}              
\def\iaucirc{IAU~Circ.}       

\usepackage{graphicx}

\shorttitle{Understanding Black Hole Formation}
\shortauthors{T. Fragos et al.}

\begin{document}

\title{	Understanding Compact Object Formation and Natal Kicks\\ 
II.\	 The case of XTE\,J1118+480} 

\author{ T.\ Fragos\altaffilmark{1}, B.\ Willems\altaffilmark{1}, V.\ Kalogera\altaffilmark{1}, N.\ Ivanova\altaffilmark{2}, G.\ Rockefeller\altaffilmark{3}, C.\ L.\ Fryer\altaffilmark{3,4}, and P.\ A.\ Young\altaffilmark{5}} 

\altaffiltext{1}{Northwestern University, Department of Physics and Astronomy, 2145 Sheridan Road, Evanston, IL 60208, USA}
\altaffiltext{2}{Canadian Institute for Theoretical Astrophysics, University of Toronto, 60 St George, Toronto, ON M5S 3H8, Canada}
\altaffiltext{3}{CCS Division, Los Alamos National Laboratory, Los Alamos, NM 87545}
\altaffiltext{4}{Department of Physics, The University of Arizona, Tucson, AZ 85721} 
\altaffiltext{5}{School of Earth and Space Exploration, Arizona State University, Tempe, AZ 85287}

\email{tassosfragos@northwestern.edu, b-willems@northwestern.edu, vicky@northwestern.edu,  nata@cita.utoronto.ca, gaber@lanl.gov, fryer@lanl.gov, patrick.young.1@asu.edu}

\begin{abstract}

In recent years, an increasing number of proper motions have been measured for Galactic X-ray binaries. When supplemented with accurate determinations of the component masses, orbital period, and donor effective temperature, these kinematical constraints harbor a wealth of information on the system's past evolution. Here, we consider all this available information to reconstruct the full evolutionary history of the black hole X-ray binary XTE\,J1118+480, assuming that the system originated in the Galactic disk and  the donor has solar metallicity. This analysis accounts for four evolutionary phases: mass transfer through the ongoing X-ray phase, tidal evolution before the onset of Roche-lobe overflow, motion through the Galactic potential after the formation of the black hole, and binary orbital dynamics due to explosive mass loss and possibly a black hole natal kick at the time of core collapse. We find that right after black hole formation, the system consists of a $\simeq 6.0 - 10.0 \, \rm M_{\odot}$ black hole and a $\simeq 1.0 - 1.6 \, \rm M_{\odot}$ main-sequence star. We also find that that an asymmetric natal kick is not only plausible but \emph{required} for the formation of this system, and derive a lower and upper limit on the black hole natal kick velocity magnitude of $80\,\rm{km\, s^{-1}}$ and $310\,\rm{km\, s^{-1}}$, respectively.

\end{abstract}


\maketitle


\section{INTRODUCTION}

In recent years the observed sample of Galactic black-hole (BH) X-ray binaries (XRBs) has increased significantly. For many of these systems there exists a wealth of observational information about their current physical state: BH and donor masses, orbital period, donor's position on the H-R diagram and surface chemical composition, transient or persistent X-ray emission, and Roche-lobe overflow (RLO) or wind-driven character of the mass-transfer (MT) process. The full proper motion has been measured for a handful of these systems \citep{Mirabel2001, Mirabel2002, Mirabel2003}, which in addition to the earlier measurements of center-of-mass radial velocities and distances gives us information about the 3-dimensional kinematic properties of these binaries. This plethora of observational results provides us with a unique opportunity to study and try to understand the formation and evolution of BHs in binaries.  This paper is the second in a series in which we address outstanding questions about the formation of compact objects in XRBs,  such as what the mass relation between compact objects and their helium star progenitors is, and whether during the core collapse, BHs receive natal kicks comparable to those of neutron stars (NS).

XTE\,J1118+480 is one of the Galactic X-ray novae that has been dynamically confirmed to contain a BH primary and the first to be currently located in the galactic halo \citep{Remillard2000, Cook2000, Uemura2000}. After the first detection several groups performed detailed observations of the binary system \citep{McClintock2001, Wagner2001, Mirabel2001, Haswell2002, Gelino2006, Gonzalez2006, Gonzalez2008}. These studies yielded strong observational constraints on the present properties of XTE\,J1118+480.

The formation and evolution of BH XRBs, in such short-period orbits ($\lesssim12\,hr$) as that of XTE\,J1118+480, is still a subject of open discussion. \citet{Haswell2002}, assuming that the accreting material in XTE\,J1118+480 has been CNO processed, showed that the system must have undergone RLO from a secondary star of about $\simeq1.5\,\rm M_{\odot}$ at a period of $\simeq15\, \rm h$. \citet{Gualandris2005} constrained the age of the system, using stellar evolution calculations, to be between 2 and 5\,Gyr, rendering a globular cluster origin unlikely. They also studied the kinematic evolution of XTE\,J1118+480, using Monte-Carlo techniques to account for the uncertainties in the observed values of the proper motion, and the distance of the binary. They inferred that the peculiar velocity of the system, acquired from the supernova (SN) explosion of the primary star, must be $183\pm31\,\rm km\,s^{-1}$ in order for the system's orbit to change from a Galactic disk orbit to the currently observed one. 

\citet{Justham2006} proposed an alternative scenario for the formation of BH XRBs with orbital periods shorter than half a day. They suggested that the companion star at the onset of the RLO, is an Ap or Bp intermediate mass star ($3\,\rm M_{\odot}\lesssim M \lesssim 5\,\rm M_{\odot}$). These types of stars are associated with  anomalously high magnetic fields. The authors suggested that the primordial magnetic field of Ap and Bp stars can lead to substantial systemic loss of angular momentum, through magnetic braking (MB), and evolve the binary to shorter orbital periods during MT. They showed that such a scenario can lead to the formation of BH binaries with population properties consistent with the observations, with exception of a discrepancy between the calculated effective temperatures and the observed spectral types of the donor stars. 

On a different approach, \citet{Yungelson2006} and \citet{Yungelson2008} studied evolutionary models for populations of short-period BH XRBs, and compared them with observations of soft X-ray transients. Assuming a population of semi-detached short-period binaries with low-mass ($\lesssim 1.5\,\rm M_{\odot}$) donors and massive ($\gtrsim 4.0 \,\rm M_{\odot}$) compact accretors, and adopting a MB rate reduced by a factor of 2 compared to the \citet{VZ1981} prescription, they showed that the calculated masses and effective temperatures of secondaries are in a satisfactory agreement with the observed ones, as inferred from their spectral types. The conclusions of this work suggest that the assumption of \citet{Justham2006} for a new kind of MB mechanism in Ap and Bp stars is not necessary for explaining the formation of BH XRBs with short-period orbits.

\citet{Ivanova2006} proposed that all short-period BH LMXB contained a pre-main-sequence (MS) secondary star at the time of the BH formation, with a subset potentially still containing a pre-MS donor at present, suggesting that the strong magnetic fields of these stars power the MB and bring the binary to contact. She showed that the orbital period and the donor's effective temperature of such X-ray binaries agree better with the available observations of BH LMXBs, than those of binaries with a MS donor, explaining simultaneously the roughly primordial abundance of Li detected in donor companions of Galactic BH LMXBs. However, this scenario is not applicable to XTE\,J1118+480, as its orbital period of $0.17\,\rm days$ is located on the left of the ZAMS line \citep[see Fig.~1 in ][]{Ivanova2006} and is explained by a MS donor star at the onset of RLO.

The works by \citet{Justham2006,Yungelson2006, Yungelson2008}, and \citet{Ivanova2006} study the population of short-period BH XRBs as a \emph{whole}, trying to propose a general formation scenario for this kind of systems. However, detailed modeling of \emph{individual} objects, which harness all the available observational data, is required in order to gain a better understanding about the properties of these systems, during the formation of the compact object.

\citet{Willems2005} showed how using all the currently available observational constraints, one can  uncover the evolutionary history of the system from the present state back to the time just prior to the core-collapse event, and they applied their analysis to the BH XRB GRO\,J1655-40. In the work presented here we follow the same framework but we focus on the case of XTE\,J1118+480. The broader scope of this project is to address open questions related to BH formation such as the relation between the masses of BHs and those of their progenitors and whether natal kicks comparable to those of NS are imparted to BHs during core collapse. 

The plan of the paper is as follows. In \S\,2 we review the currently available constraints on the properties of the Galactic BH XRB XTE\,J1118+480. A general outline of the analysis used to reconstruct the system's evolutionary history is presented in \S\,3, while the individual steps of the analysis and the resulting constraints on the formation of the BH are discussed in \S\,4. In \S\,5, hydrodynamic core-collapse simulations are presented and the nature of the BH progenitor is constrained even further. In \S,6, we discuss some of the assumptions introduced in our analysis and comment on possible alternative formation scenarios for XTE\,J1118+480. The final section is devoted to a summary of our results and some concluding remarks.

\section{OBSERVATIONAL CONSTRAINTS FOR XTE\,J1118+480}

XTE\,J1118+480, first detected with the Rossi X-Ray Timing Explorer All-Sky Monitor \citep{Remillard2000}, is the $11^{th}$ X-ray novae that has been dynamically confirmed to contain a BH primary and the first to be located at a high galactic latitude ($l=157.7^o$, $b=+62.3^o$). \citet{Cook2000} and \citet{Uemura2000} soon reported that the light curve of the optical counterpart was smooth and sinusoidal, showing a regular periodic variation with a period of $0.1706\pm 0.0009\, \rm{d}$  and a full amplitude of $0.061\pm 0.0005\, \rm{mag}$. 

Observations with the 6.5\,m Multiple Mirror Telescope by \citet{McClintock2001} revealed that the amplitude of the secondary star's radial velocity curve is $698\pm 14\, \rm{km\, s^{-1}}$. The period was more accurately determined to be $0.17013\pm 0.00010\, \rm{d}$ and the mass function $f(M)=6.00\pm0.36\,\rm{M_\odot}$, implying a minimum mass of $6\, \rm{M_\odot}$ for the compact object. Assuming an ellipsoidal light curve with no accretion disk contribution they derived a maximum BH mass of $10\, \rm{M_\odot}$. Finally, they estimated that the spectral type of the secondary star ranges from $K5\, V$ to $M1\, V$ and, using this range, derived a distance of $1.8\pm 0.6\, \rm{kpc}$. \citet{Wagner2001}, using a combination of data from the $6.5\, \rm{m}$ Multiple Mirror Telescope, the 4.2m William Herschel Telescope and the Instituto de Astrof\'isica de Canarias 0.8\,m telescope, put some tighter constraints on the properties of the system. They derived an  orbital period of $P=0.169930\pm 0.000004\, \rm{d}$ and  a mass function of $6.1\pm 0.3\, \rm{M_\odot}$. They estimated the spectral type of the secondary to be $K7\, V - M0\, V$, and from the modeling of the light curve derived a high orbital inclination $i=81^o \pm 2^o$ and a BH mass in the range of $M_{BH}=6.0-7.7 \, \rm{M_\odot}$ (90\% confidence), for plausible secondary star masses of $M_2=0.09-0.5\, M_\odot$. Their estimate for the distance was $d=1.9 \pm 0.4\, \rm{kpc}$.

The proper motion of XTE\,J1118+480 was observed with the VLBA on 4 May -- 24 July 2000 \citep{Mirabel2001}. During this period the system's position shifted at a rate of $-16.8\pm 1.6\, \rm{mas\, yr^{-1}}$ in right ascension and $-7.4 \pm 1.6\, \rm{mas\, yr^{-1}}$ in declination. In order to derive 3-D velocity components, the authors described the motion of the system with respect to a right-handed Cartesian frame of reference $OXYZ$ whose origin coincides with the Galactic center and whose $XY$-plane coincides with the mid-plane of the Galactic disk. The direction from the projection of the Sun's position onto the Galactic plane to the Galactic center is taken as the positive direction of the $X$-axis, the direction from the Sun to the Northern Galactic pole as the positive direction of the $Z$-axis, and the direction of the Galactic rotational velocity at the position of the Sun as the positive direction of the $Y$-axis. In terms of these coordinates, the velocity components U, V and W with respect to the $X$-, $Y$-, and $Z$-axes, were calculated to be $U=-105 \pm 16\, \rm{km\, s^{-1}}$, $V=-98 \pm 16\, \rm{km\, s^{-1}}$  and $W=-21 \pm 10\, \rm{km\, s^{-1}}$, taking into account the mean values and error estimates from the different measurements for the distance, radial velocity, and proper motion. \citet{Mirabel2001} concluded that only an extraordinary kick from the SN explosion could have launched the black hole into the observed Galactic halo orbit from a birthplace in the disk of the Galaxy. The authors therefore favored a globular cluster origin of this system as a more probable scenario. 

\citet{Haswell2002} performed ultraviolet spectroscopy of  XTE\,J1118+480. The carbon and oxygen lines were undetectable, while the nitrogen emission appeared enhanced. However, they were not able to derive quantitative limits on the surface mass ratio of these elements. They concluded that the emission line strengths in this system strongly suggest that the accreting material has been significantly CNO processed. The spectrum therefore implies that the companion star in XTE\,J1118+480 must be partially nuclearly evolved and have lost its outer layers, exposing inner layers which have been mixed with CNO processed material from the central nuclear region. Therefore, they concluded that the MT must have been initiated from a somewhat evolved and sufficiently massive ($\sim 1.5\,\rm M_{\odot}$) donor.

\citet{Gelino2006} using the 1.5 m telescope at the TUBITAK National Observatory, obtained optical and infrared photometry of XTE\,J1118+480 in its quiescent state. Their results yield somewhat different values for the properties of the system. Comparing the observed spectra with synthetic ones, they reported that the most likely spectral type of the donor star is $K7\, V$, while the most likely value for the orbital inclination angle is $68^o\pm2^o$. This inclination angle corresponds to a primary BH mass of $8.53\pm 0.60\, \rm{M_\odot}$, and a distance of $1.72\pm 0.10\, \rm{kpc}$. More recent observations with the 10-m KECK II telescope, equipped with the Echellette Spectrograph and Imager \citep{Gonzalez2006}, revealed that the secondary star has a super-solar surface metallicity $[Fe/H]=0.2 \pm 0.2$. A galactic halo origin of the system cannot be excluded by this result since, as they explain, the high metallicity might be due to pollution of the secondary star's surface during the SN event. In a follow-up investigation, \citet{Gonzalez2008} used a grid of SN explosion models with different He core masses, metallicities, and geometries of the SN explosion to explore the possible formation scenarios for XTE\,J1118+480. They found that metal-poor models associated with a formation scenario in the Galactic halo provide unacceptable fits to the observed abundances, allowing them to reject a halo origin for this XRB. The thick-disk scenario produced better fits, although they required substantial fallback and very efficient mixing processes, making  a thick disk origin quite unlikely as well. The best agreement between the model predictions and the observed abundances was obtained for metal-rich progenitor models, corresponding to a birth of the system in the thin Galactic disk.

In our analysis, we use the most recent observational constraints for the various system parameters described in the previous paragraphs. For convenience, the adopted constraints are summarized in Table~\ref{1118param}.

\begin{deluxetable}{lccc}
\tablecolumns{4}
\tabletypesize{\scriptsize}
\tablecaption{Properties of XTE\,J1118-480. 
\label{1118param}} 
\tablehead{ \colhead{Parameter} & 
     \colhead{Notation} & 
     \colhead{Value} &
	 \colhead{References\tablenotemark{a}}
	}
\startdata
Distance															& $d$			& $1.85 \pm 0.36$\,kpc 				& (2), (3), (4) \\
Galactic longitude													& $l$           & $157.7^\circ$ 					&  (1) \\
Galactic latitude													& $b$           & $+62.3^\circ$  			 		&  (1) \\
Velocity towards the Galactic Center\tablenotemark{b}				& $U$			& $-105 \pm 16$\,km\,s$^{-1}$ 		& (4) \\
Velocity in the direction of the Galactic Rotation\tablenotemark{b}	& $V$			& $-98 \pm 16$\,km\,s$^{-1}$   		& (4) \\
Velocity towards the Northern Galactic Pole\tablenotemark{b}		& $W$			& $-21 \pm 10$\,km\,s$^{-1}$  		& (4) \\
Orbital Period														& $P_{\rm orb}$	& $0.17 \pm 0.001$\,days 			& (2), (3) \\
BH Mass																& $M_{\rm BH}$	& $8.0 \pm 2.0\,\rm{M_\odot}$  		& (2), (3), (5) \\
Donor Mass															& $M_2$       	& $0.455 \pm 0.355\,\rm{M_\odot}$  	& (2), (3), (5) \\
Donor Luminosity													& $L_2$       	& $0.0594 \pm 0.0319\,\rm{L_\odot}$ & (2), (3) \\
Donor Effective Temperature											& $T_{\rm eff2}$& $4409 \pm 440$\,K       			&  (2), (3) \\
Donor Surface Metallicity											& $[Fe/H]$    	& $0.2 \pm 0.2$           			&  (6) \\
\enddata
\tablenotetext{a}{(1) \citet{Remillard2000}, (2) \citet{McClintock2001}, (3) \citet{Wagner2001}, (4) \citet{Mirabel2001}, (5) \citet{Gelino2006}, (6) \citet{Gonzalez2006}}
\tablenotetext{b}{relative to the local standard of rest}
\end{deluxetable}

\section{OUTLINE OF ANALYSIS METHODOLOGY}

In our analysis, we assume that XTE\,J1118+480 formed in the Galactic disk from the evolution of an isolated primordial binary consisting of solar metallicity stars. A possible globular cluster origin of the system is discussed in \S\ref{GC}. 

The standard formation channel \citep[e.g. ][]{BvdH1991,TvdH2006} of BH XRBs in the Galactic field involves  a primordial binary with a large mass ratio and a primary more massive than $20-25\,\rm{M_\odot}$ so that the formation of a BH is possible. The more massive star evolves quickly  to the giant branch and, if the initial orbital period is less that $\sim 10\,\rm yr$, the primary soon overflows its Roche-lobe and initiates an unstable common envelope phase. During this  phase, the less massive secondary, which is still relatively unevolved, orbits inside the envelope of the primary and is assumed to remain intact. The orbit of the system changes dramatically  though, as orbital energy is lost due to friction between the secondary star and the envelope of the giant. Part of the lost orbital energy is used to expel the hydrogen-rich envelope of the primary star.  If the energy dissipated is enough to expel the whole envelope and prevent a binary merger, the common envelope phase results in a binary system consisting of a relatively unevolved low-mass MS star orbiting around the naked helium core of the primary star in a short-period orbit. The massive helium core soon reaches core collapse to form a compact object at which time the binary orbit is altered due to mass loss and possibly an natal kick imparted to the compact object. If the binary survives all these stages, angular momentum loss mechanisms, such as MB, tides, and gravitational wave radiation, will shrink the orbit further, while the low-mass companion may evolve off the MS. Consequently, the companion star eventually overflows its Roche-lobe, transferring mass onto the compact object and initiating an XRB phase.

The analysis we follow in order to trace back the evolutionary history of XTE\,J1118+480 incorporates a number of calculations which can be summarized in four discrete steps.

\begin{figure}[ht]
\plotone{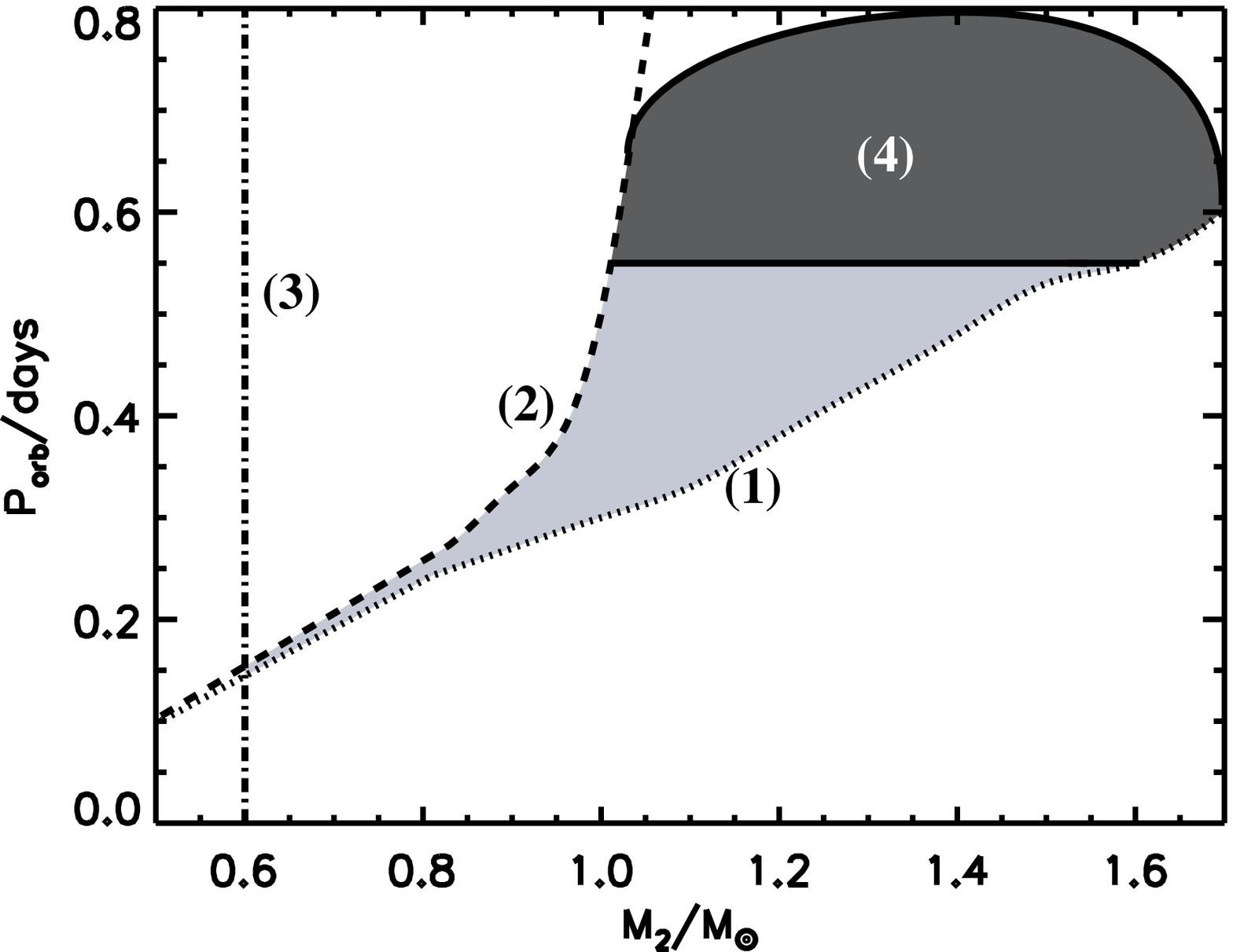}
\caption{Constraints on the XTE\,J1118+480 parameter space at the onset of RLO. A minimum initial period is set by requiring that the Roche-lobe radius at the onset of the MT has to be larger than the ZAMS radius of the donor (dotted line 1), while a  maximum initial period is set by requiring that the donor fills its Roche lobe within a Hubble time (dashed line 2). At the onset of RLO, the secondary star should furthermore be more massive than the currently observed donor (dashed-dotted line 3). The period at the onset of RLO must also be below the bifurcation period, which depends on the adopted MB prescription. The range of bifurcation periods covered by the different MB prescriptions is represented by the dark grey shaded region (4). This figure is calculated assuming an initial BH mass of $8.0\,\rm M_{\odot}$. However, the shape of the allowed parameter space is rather insensitive to the mass of the BH.}
\label{ParSp}
\end{figure}

We first use a binary stellar evolution code \citep{Podsiadlowski2002,Ivanova2003,Kalogera2004} and calculate a grid of evolutionary sequences for binaries in which a BH is accreting mass from a Roche-lobe filling companion. The parameter space (initial donor mass, BH mass, and orbital period at the start of RLO) that we have to cover with this grid is finite and limited by a number of constraints visualized in Fig.~\ref{ParSp}. First, the initial Roche-lobe radius at the onset of MT has to be larger than the ZAMS radius of the donor. This sets a minimum initial period for given masses of the two components (dotted line 1). A maximum initial period is set by requiring that the donor fills its Roche lobe within a Hubble time (dashed line 2). Moreover, a minimum mass for the donor star is set by the fact that the secondary star, at the onset of RLO, must be more massive than the currently observed donor star (dashed-dotted line 3). Finally, given that the observed period is below the bifurcation period\footnote{The bifurcation period is the critical orbital period that separates the formation of converging systems (which evolve towards shorter orbital period until the mass-losing component becomes degenerate and an ultra-compact binary is formed) from the formation of diverging systems (which  evolve towards longer orbital period until the mass-losing star has lost its envelope and a wide detached binary is formed).}, the period at the onset of RLO should also be below. The value of the bifurcation period depends on the masses of the system and the adopted MB prescription (dark grey region 4). For the initial mass of the BH we use a range of $6.0-8.0\,\rm M_{\odot}$.

To explore the full range of possible MT sequences, we calculate sequences for both conservative and fully non-conservative MT. In the case of conservative MT, the MT rate is still limited to the Eddington rate. Any matter transfered in excess of the Eddington rate is assumed to escape from the system carrying away the specific angular momentum of the accretor. We use the same assumption for fully non-conservative MT, where all the transfered matter escapes from the system. For each sequence, we examine whether at any point in time the calculated binary properties are in agreement with all available observational measurements simultaneously. From the ``successful'' sequences we derive the properties of the binary at the onset of the RLO phase: initial BH and donor masses, orbital period, and age of the donor star. The time at which the successful sequences satisfy all observational constraints furthermore provides an estimate for the donor's current age.

Next, we consider the kinematic evolutionary history of the XRB in the Galactic potential. In particular, we use the current position and the measured 3D velocity with their associated uncertainties to trace the Galactic motion back in time. Combined with the tight constraints on the current age of the system given by the successful MT sequences, this allows us to determine the position and velocity of the binary at the time of BH formation (we refer to these as the ``birth'' position and velocity). We use this birth position and velocity to estimate the {\em peculiar} velocity of the binary right after the formation of the BH by subtracting the local Galactic rotational velocity at the birth position from the system's total center-of-mass velocity.

In the third step, we follow the binary orbital evolution of the system due to tides, MB, and gravitational radiation between the time of BH formation and the onset of the XRB phase. This calculation yields the post-collapse semi-major axis and eccentricity of the system, back from the start of RLO to the moment of core collapse. 

Finally, knowing all the properties of the system just after the BH formation, we analyze the orbital dynamics of the core collapse event due to mass loss and possible natal kicks imparted to the BH. In what follows, we refer to the instants just before and right after the formation of the BH by the terms pre-SN and post-SN, respectively. If the BH is formed via an intermediary NS stage followed by fall back of some fraction of the SN material, these two times may be slightly offset from each other. We here neglect this small offset and assume the entire SN process to be instantaneous regardless of the formation mechanism of the BH. Starting from this assumption and the constraints derived for the post-SN binary properties, we use energy and angular momentum conservation during the core collapse event to constrain the pre-SN binary properties (BH progenitor mass and orbital separation) and the natal kick (magnitude and direction) that may have been imparted to the BH.

\section{RESULTS}

\subsection{Mass-Transfer Calculations}

We calculated a grid of 792 evolutionary sequences for mass-transferring BH XRBs covering the available initial parameter space for the masses of the BH and the donor star, and the orbital period at the onset of the RLO (see Fig.~\ref{ParSp}). Since we assume that the system was formed from an isolated primordial binary in the Galactic disk, we adopt a solar metallicity for the donor star in our MT calculations.

As mentioned earlier, in order to be conservative with our assumptions, we tried to take into account all possible uncertainties affecting the MT phase. Hence, we considered both conservative (but Eddington limited) and fully non-conservative MT as a lower and upper limit to the angular momentum loss due to mass loss from the system. We also adopted two prescriptions for the angular momentum loss rate due to MB for late-type stars: the \citet{IT2003} prescription which is based on an idealized two-component coronal model and takes into account recent observational data, and the ``standard'' prescription by \citet{RVJ1983} which results in much stronger angular momentum losses. In addition we used two different criteria for the application of MB. In the first case we applied MB to all systems in which the donor's mass is below $1.5\,M_{\odot}$, while in the second case, we applied MB only when the donor has an outer convective envelope and an inner radiative core (relevant to stars with mass less than $\sim 1.3\,M_{\odot}$).

\begin{figure}[ht]
\plottwo{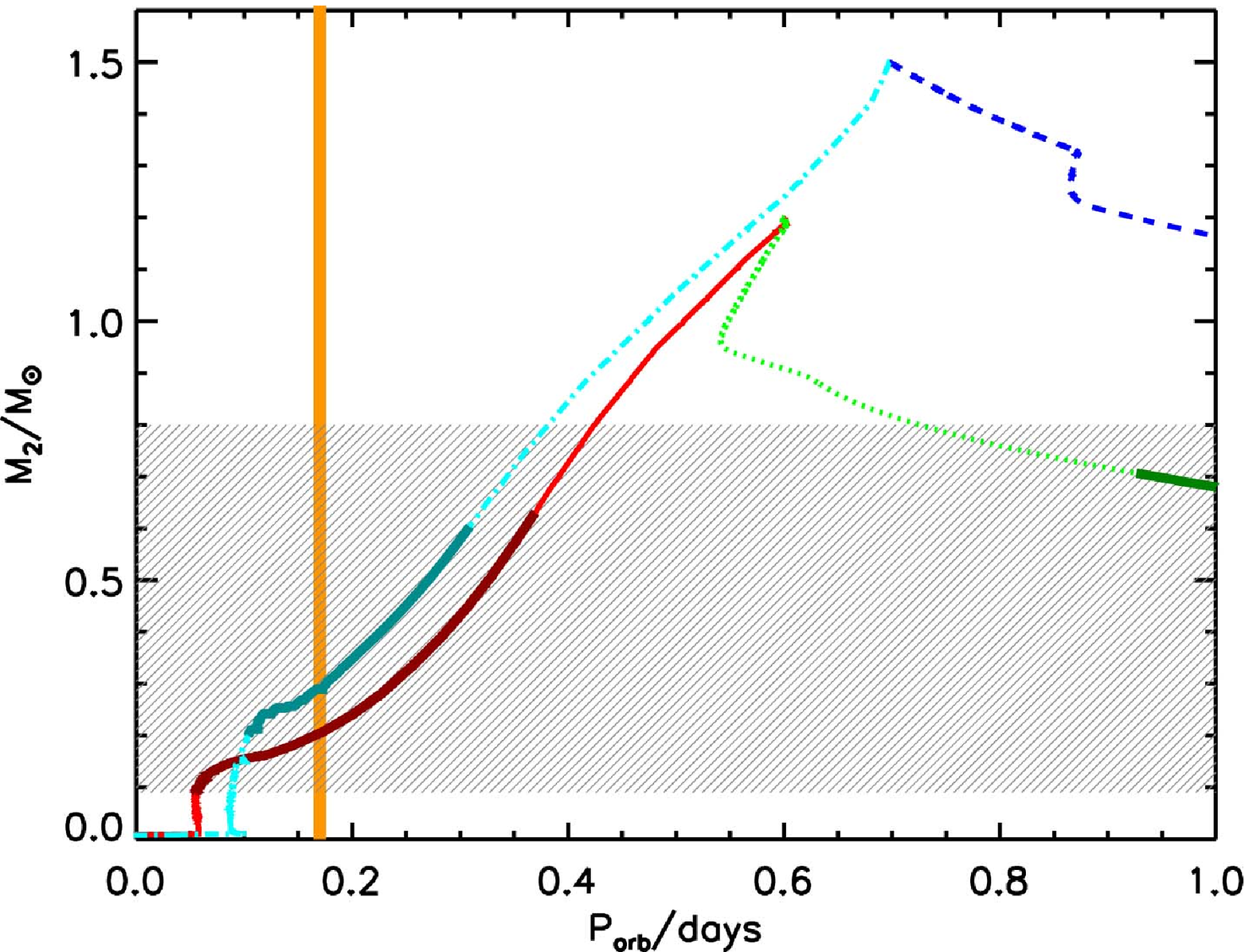}{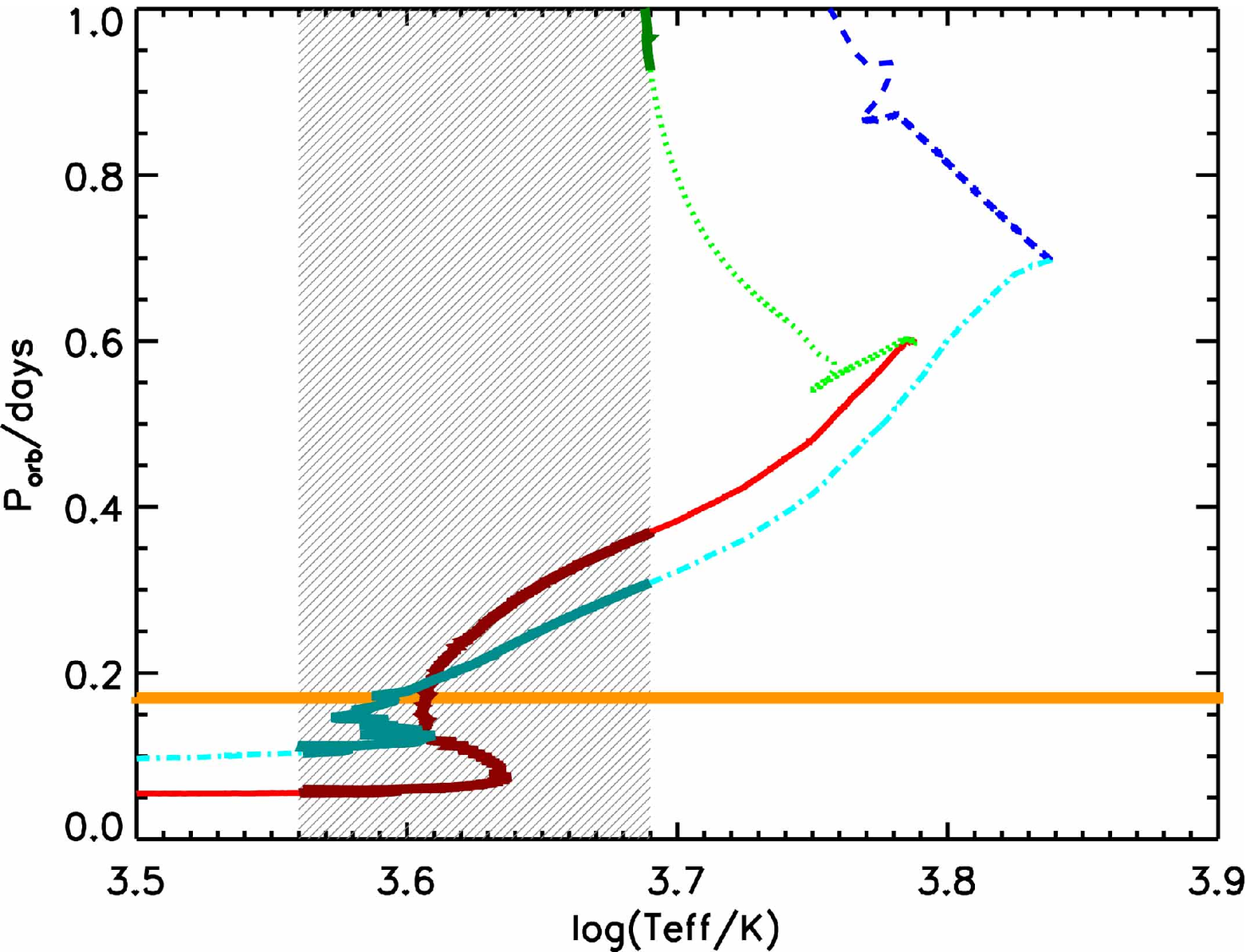}
\caption{Systematic behavior of some selected MT sequences with different initial component masses and orbital periods. On the \emph{left panel}, the variation of the donor mass is displayed as a function of the orbital period. The grey hatched region indicates the observational constraint on the present-day donor mass. The \emph{right panel} shows the variation of the orbital period as a function of the donor star's effective temperature. The grey hatched region indicates the observational constraint on the effective temperature of the donor star. The  \emph{vertical line} in the $(M_{\rm 2},P_{\rm orb})$-plot and the \emph{horizontal line} in the $(P_{\rm orb},T_{eff})$-plot represents the observed orbital period of 0.167\,days. On both panels, the thick part of the evolutionary tracks indicates the part of the sequence where the donor's effective temperature satisfies the observational temperature constraint. Successful sequences able to simultaneously satisfy all observational constraints are therefore given by tracks for which the thick part crosses the $P_{\rm orb}$ constraint line inside the hatched region on the left panel. On both panels the \emph{solid line} corresponds to a MT sequence with $M_2=1.2\,\rm M_{\odot}$, $M_{\rm BH}=6.0\,\rm M_{\odot}$, and $P_{\rm orb}=0.6\,\rm days$ at the start of RLO, and with the \citet{RVJ1983} MB rate applied to stars consisting of a convective envelope and a radiative core. The \emph{dotted line} corresponds to a MT sequence with the same initial binary properties, but with the \citet{IT2003} MB prescription applied to stars consisting of a convective envelope and a radiative core. The \emph{dashed line} indicates a MT sequence with $M_2=1.5\rm\,M_{\odot}$, $M_{\rm BH}=8.0\,\rm M_{\odot}$, and $P_{\rm orb}=0.7\,\rm days$ at the start of RLO, and the \citet{RVJ1983} MB rate applied to stars consisting of a convective envelope and a radiative core, while the \emph{dash-dotted line} corresponds to a MT sequence with the same properties, but with the  \citet{RVJ1983}  MB applied to all stars with mass below $1.5\,M_{\odot}$, even if they do not have a convective envelope. In all four cases, MT is assumed to be fully non-conservative.}
\label{MTseq}
\end{figure}

For each of the calculated MT sequences, we checked whether there is a time during the evolution of the system at which its properties satisfy simultaneously all the observational constraints: the masses of the BH and the donor star, the temperature of the donor star, the orbital period. Since XTE\,J1118+480 is a soft X-ray transient, we also check whether the MT rate is below the critical rate ($\dot{M}_{\rm crit}$) for the occurrence of thermal disk instabilities, believed to cause the transient behavior of low-mass XRBs \citep{Paradijs1996, KKB1996, DLHC1999, MPH2002}. In Figure \ref{MTseq} we show, as an example, the systematic behavior of some selected MT sequences with different initial component masses, orbital periods, and MB prescriptions.  Starting with $M_2=1.2\,M_{\odot}$, $M_{\rm BH}=6.0\,M_{\odot}$ and $P_{\rm orb}=0.6$\,days at the onset of RLO, and assuming non-conservative MT, the weaker \citet{IT2003} MB law (dotted line) does not lead to strong enough angular momentum losses to shrink the orbit down to the observed period of $0.17 \pm 0.001$\,days. Instead, the nuclear evolution of the donor dominates the orbital evolution during MT, resulting in the widening of the orbit. Starting from the same initial properties but adopting the \citet{RVJ1983} MB law (solid line) leads to a decrease of the orbital period which eventually reaches the currently observed period, simultaneously satisfying all the other observational constraints. Adopting the  \citet{RVJ1983} MB law to a binary with a more massive donor star and a wider orbit at the onset of RLO ($M_2=1.5\,M_{\odot}$, $M_{BH}=8.0\,M_{\odot}$ and $P_{orb}=0.7\,days$) has a similarly prominent effect. In particular, such a donor star has no outer convective envelope, so that no angular momentum loss mechanism operates when MB is applied exclusively to stars 
with a convective envelope. The orbit therefore quickly expands due to the nuclear evolution of the donor star (dashed line). On the other hand, when MB is applied to all stars with mass below $1.5\,M_{\odot}$, the orbit contracts and at some point the system is able to simultaneously satisfy all the observational constraints  (dash-dotted line).

\begin{figure}[ht]
\plotone{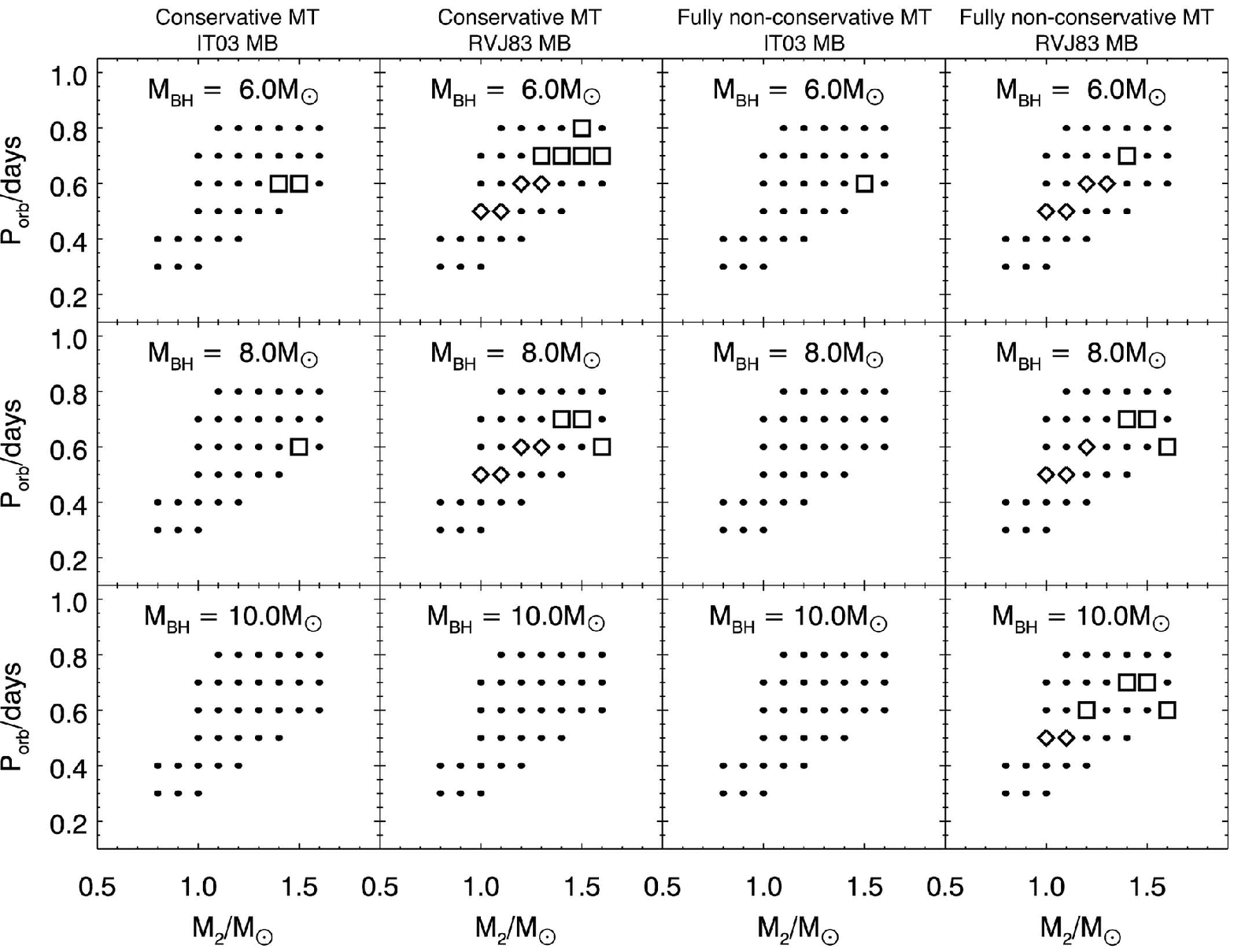}
\caption{Initial donor masses and orbital periods of the evolutionary sequences at RLO onset.  \emph{Solid circles} indicate MT sequences that fail to simultaneously satisfy all the current observational constraints. \emph{Diamonds} indicate MT sequences which are able to simultaneously satisfy all the current observational constraints assuming that MB operates only when a convective envelope and a radiative core are present in the donor star. \emph{Squares} correspond to the \emph{additional} MT sequences that satisfy the observational constraints when we assume that MB operates to all stars with mass below $1.5\,M_{\odot}$. In the labels of the figure, IT03 indicates that the \citet{IT2003} MB rate was applied, while RVJ83 corresponds to the \citet{RVJ1983} MB prescription.}
\label{MPI}
\end{figure}

\begin{deluxetable}{lcccccccccccccccc}
\tablecolumns{17}
\tabletypesize{\scriptsize}
\rotate
\tablewidth{585pt}
\tablecaption{Selected properties of MT sequences calculated to satisfy the observational constraints for XTE\,1118+480. The current parameters correspond to the point where the binary's orbital period is equal to the observed orbital period of 0.17\,days.    
\label{sucMT}}
\tablehead{ 
   \colhead{} & 
   \multicolumn{5}{c}{Parameters at onset of RLO} &
   \colhead{} &
   \multicolumn{6}{c}{Current parameters}  &
   \colhead{} &
   \multicolumn{3}{c}{MT parameters}  \\
   \cline{2-6} \cline{8-13}  \cline{15-17} \\
   \colhead{Sequence} & 
   \colhead{$M_{\rm BH}$\tablenotemark{a}} & 
   \colhead{$M_2$\tablenotemark{b}} & 
   \colhead{$P_{\rm orb}$\tablenotemark{c}} &
   \colhead{$X_2$\tablenotemark{d}} &
   \colhead{$\tau_2$\tablenotemark{e}} &
   \colhead{} &
   \colhead{$M_{\rm BH}$} & 
   \colhead{$M_2$} & 
   \colhead{$\log{(L_2)}$\tablenotemark{f}} &
   \colhead{$\log{(T_{\rm eff2})}$\tablenotemark{g}} &
   \colhead{$X_2$} &
   \colhead{$\tau_2$} &
   \colhead{} &
   \colhead{MT type\tablenotemark{h}} &
   \colhead{MB rate\tablenotemark{i}} &
   \colhead{MB criterion\tablenotemark{j}} \\
   \colhead{} & 
   \colhead{($M_\odot$)} & 
   \colhead{($M_\odot$)} & 
   \colhead{(days)} &
   \colhead{} &
   \colhead{(Gyr)} &
   \colhead{} &
   \colhead{($M_\odot$)} & 
   \colhead{($M_\odot$)} & 
   \colhead{($L_\odot$)} &
   \colhead{(K)} &
   \colhead{} &
   \colhead{(Myr)} &
   \colhead{} &
   \colhead{} &
   \colhead{} &
   \colhead{}
   }
\startdata
\multicolumn{17}{l}{} \\
\newcounter{Model}
\stepcounter{Model}Model \arabic{Model} &  6.0 & 1.0 & 0.5 & 0.00 & 10.08 & &  6.79 & 0.21 & -1.47 &  3.62 & 0.00 & 11.42 & &  cons. MT  &  RVJ83  &  conv. env.  \\ 
\stepcounter{Model}Model \arabic{Model} &  8.0 & 1.0 & 0.5 & 0.00 & 10.10 & &  8.80 & 0.20 & -1.45 &  3.62 & 0.00 & 11.77 & &  cons. MT  &  RVJ83  &  conv. env.  \\ 
\stepcounter{Model}Model \arabic{Model} &  6.0 & 1.1 & 0.5 & 0.00 &  6.00 & &  6.85 & 0.25 & -1.59 &  3.58 & 0.00 &  7.32 & &  cons. MT  &  RVJ83  &  conv. env.  \\ 
\stepcounter{Model}Model \arabic{Model} &  8.0 & 1.1 & 0.5 & 0.00 &  6.02 & &  8.86 & 0.24 & -1.60 &  3.58 & 0.00 &  7.55 & &  cons. MT  &  RVJ83  &  conv. env.  \\ 
\stepcounter{Model}Model \arabic{Model} &  6.0 & 1.2 & 0.6 & 0.19 &  4.87 & &  6.99 & 0.21 & -1.52 &  3.61 & 0.06 &  6.66 & &  cons. MT  &  RVJ83  &  conv. env.  \\ 
\stepcounter{Model}Model \arabic{Model} &  8.0 & 1.2 & 0.6 & 0.19 &  4.87 & &  9.00 & 0.20 & -1.46 &  3.62 & 0.03 &  7.05 & &  cons. MT  &  RVJ83  &  conv. env.  \\ 
\stepcounter{Model}Model \arabic{Model} &  6.0 & 1.3 & 0.6 & 0.35 &  2.88 & &  7.09 & 0.21 & -1.40 &  3.63 & 0.05 &  5.77 & &  cons. MT  &  RVJ83  &  conv. env.  \\ 
\stepcounter{Model}Model \arabic{Model} &  8.0 & 1.3 & 0.6 & 0.35 &  2.89 & &  9.10 & 0.20 & -1.34 &  3.65 & 0.02 &  6.20 & &  cons. MT  &  RVJ83  &  conv. env.  \\ 
\stepcounter{Model}Model \arabic{Model} & 10.0 & 1.0 & 0.5 & 0.00 & 10.13 & & 10.00 & 0.20 & -1.41 &  3.63 & 0.00 & 12.10 & &  non-cons. MT  &  RVJ83  &  conv. env.  \\ 
\stepcounter{Model}Model \arabic{Model} &  6.0 & 1.0 & 0.5 & 0.00 & 10.08 & &  6.00 & 0.21 & -1.47 &  3.62 & 0.00 & 11.45 & &  non-cons. MT  &  RVJ83  &  conv. env.  \\ 
\stepcounter{Model}Model \arabic{Model} &  8.0 & 1.0 & 0.5 & 0.00 & 10.10 & &  8.00 & 0.20 & -1.44 &  3.62 & 0.00 & 11.79 & &  non-cons. MT  &  RVJ83  &  conv. env.  \\ 
\stepcounter{Model}Model \arabic{Model} & 10.0 & 1.1 & 0.5 & 0.00 &  6.05 & & 10.00 & 0.24 & -1.56 &  3.58 & 0.00 &  7.77 & &  non-cons. MT  &  RVJ83  &  conv. env.  \\ 
\stepcounter{Model}Model \arabic{Model} &  6.0 & 1.1 & 0.5 & 0.00 &  6.00 & &  6.00 & 0.23 & -1.63 &  3.57 & 0.00 &  7.38 & &  non-cons. MT  &  RVJ83  &  conv. env.  \\ 
\stepcounter{Model}Model \arabic{Model} &  8.0 & 1.1 & 0.5 & 0.00 &  6.02 & &  8.00 & 0.24 & -1.59 &  3.58 & 0.00 &  7.59 & &  non-cons. MT  &  RVJ83  &  conv. env.  \\ 
\stepcounter{Model}Model \arabic{Model} &  6.0 & 1.2 & 0.6 & 0.19 &  4.87 & &  6.00 & 0.21 & -1.51 &  3.61 & 0.05 &  6.69 & &  non-cons. MT  &  RVJ83  &  conv. env.  \\ 
\stepcounter{Model}Model \arabic{Model} &  8.0 & 1.2 & 0.6 & 0.19 &  4.87 & &  8.00 & 0.20 & -1.44 &  3.63 & 0.02 &  7.12 & &  non-cons. MT  &  RVJ83  &  conv. env.  \\ 
\stepcounter{Model}Model \arabic{Model} &  6.0 & 1.3 & 0.6 & 0.35 &  2.88 & &  6.00 & 0.20 & -1.36 &  3.65 & 0.02 &  5.91 & &  non-cons. MT  &  RVJ83  &  conv. env.  \\ 
\stepcounter{Model}Model \arabic{Model} &  6.0 & 1.4 & 0.6 & 0.47 &  1.63 & &  7.14 & 0.26 & -1.49 &  3.60 & 0.13 &  7.82 & &  cons. MT  &  IT03  &  $M_{2} < 1.5 \,\rm M_{\odot}$  \\ 
\stepcounter{Model}Model \arabic{Model} &  6.0 & 1.5 & 0.6 & 0.56 &  0.90 & &  7.21 & 0.29 & -1.54 &  3.58 & 0.22 &  6.59 & &  cons. MT  &  IT03  &  $M_{2} < 1.5 \,\rm M_{\odot}$  \\ 
\stepcounter{Model}Model \arabic{Model} &  8.0 & 1.5 & 0.6 & 0.56 &  0.90 & &  9.23 & 0.27 & -1.42 &  3.61 & 0.15 &  7.39 & &  cons. MT  &  IT03  &  $M_{2} < 1.5 \,\rm M_{\odot}$  \\ 
\stepcounter{Model}Model \arabic{Model} &  6.0 & 1.4 & 0.7 & 0.34 &  2.36 & &  7.13 & 0.27 & -1.46 &  3.60 & 0.20 &  3.97 & &  cons. MT  &  RVJ83  &  $M_{2} < 1.5 \,\rm M_{\odot}$  \\ 
\stepcounter{Model}Model \arabic{Model} &  8.0 & 1.4 & 0.7 & 0.34 &  2.36 & &  9.14 & 0.26 & -1.42 &  3.61 & 0.16 &  4.43 & &  cons. MT  &  RVJ83  &  $M_{2} < 1.5 \,\rm M_{\odot}$  \\ 
\stepcounter{Model}Model \arabic{Model} &  6.0 & 1.5 & 0.7 & 0.43 &  1.52 & &  7.19 & 0.31 & -1.50 &  3.58 & 0.30 &  2.95 & &  cons. MT  &  RVJ83  &  $M_{2} < 1.5 \,\rm M_{\odot}$  \\ 
\stepcounter{Model}Model \arabic{Model} &  6.0 & 1.5 & 0.8 & 0.33 &  1.92 & &  7.25 & 0.25 & -1.13 &  3.69 & 0.07 &  4.51 & &  cons. MT  &  RVJ83  &  $M_{2} < 1.5 \,\rm M_{\odot}$  \\ 
\stepcounter{Model}Model \arabic{Model} &  8.0 & 1.5 & 0.7 & 0.43 &  1.52 & &  9.20 & 0.30 & -1.46 &  3.59 & 0.27 &  3.32 & &  cons. MT  &  RVJ83  &  $M_{2} < 1.5 \,\rm M_{\odot}$  \\ 
\stepcounter{Model}Model \arabic{Model} &  6.0 & 1.6 & 0.7 & 0.48 &  1.05 & &  7.32 & 0.28 & -1.20 &  3.66 & 0.16 &  3.65 & &  cons. MT  &  RVJ83  &  $M_{2} < 1.5 \,\rm M_{\odot}$  \\ 
\stepcounter{Model}Model \arabic{Model} &  8.0 & 1.6 & 0.6 & 0.59 &  0.56 & &  9.25 & 0.35 & -1.53 &  3.56 & 0.38 &  2.40 & &  cons. MT  &  RVJ83  &  $M_{2} < 1.5 \,\rm M_{\odot}$  \\ 
\stepcounter{Model}Model \arabic{Model} &  6.0 & 1.5 & 0.6 & 0.56 &  0.90 & &  6.00 & 0.28 & -1.47 &  3.60 & 0.18 &  7.10 & &  non-cons. MT  &  IT03  &  $M_{2} < 1.5 \,\rm M_{\odot}$  \\ 
\stepcounter{Model}Model \arabic{Model} & 10.0 & 1.4 & 0.7 & 0.34 &  2.37 & & 10.00 & 0.24 & -1.36 &  3.63 & 0.10 &  5.01 & &  non-cons. MT  &  RVJ83  &  $M_{2} < 1.5 \,\rm M_{\odot}$  \\ 
\stepcounter{Model}Model \arabic{Model} &  6.0 & 1.4 & 0.7 & 0.34 &  2.36 & &  6.00 & 0.26 & -1.46 &  3.60 & 0.19 &  4.02 & &  non-cons. MT  &  RVJ83  &  $M_{2} < 1.5 \,\rm M_{\odot}$  \\ 
\stepcounter{Model}Model \arabic{Model} &  8.0 & 1.4 & 0.7 & 0.34 &  2.36 & &  8.00 & 0.25 & -1.41 &  3.62 & 0.15 &  4.50 & &  non-cons. MT  &  RVJ83  &  $M_{2} < 1.5 \,\rm M_{\odot}$  \\ 
\stepcounter{Model}Model \arabic{Model} & 10.0 & 1.5 & 0.7 & 0.43 &  1.53 & & 10.00 & 0.28 & -1.41 &  3.61 & 0.22 &  3.79 & &  non-cons. MT  &  RVJ83  &  $M_{2} < 1.5 \,\rm M_{\odot}$  \\ 
\stepcounter{Model}Model \arabic{Model} &  8.0 & 1.5 & 0.7 & 0.43 &  1.52 & &  8.00 & 0.29 & -1.47 &  3.60 & 0.26 &  3.40 & &  non-cons. MT  &  RVJ83  &  $M_{2} < 1.5 \,\rm M_{\odot}$  \\ 
\stepcounter{Model}Model \arabic{Model} & 10.0 & 1.6 & 0.6 & 0.59 &  0.57 & & 10.00 & 0.33 & -1.50 &  3.58 & 0.34 &  2.76 & &  non-cons. MT  &  RVJ83  &  $M_{2} < 1.5 \,\rm M_{\odot}$  \\ 
\stepcounter{Model}Model \arabic{Model} &  8.0 & 1.6 & 0.6 & 0.59 &  0.56 & &  8.00 & 0.34 & -1.51 &  3.57 & 0.36 &  2.56 & &  non-cons. MT  &  RVJ83  &  $M_{2} < 1.5 \,\rm M_{\odot}$  \\ 
\enddata
\tablenotetext{a}{BH mass}
\tablenotetext{b}{Donor mass}
\tablenotetext{c}{Orbital period}
\tablenotetext{d}{Central hydrogen fraction}
\tablenotetext{e}{Age}
\tablenotetext{f}{Donor luminosity}
\tablenotetext{g}{Donor effective temperature}
\tablenotetext{h}{MT type: ``cons. MT'' denotes conservative but Eddington limited MT, while ``non-cons. MT'' corresponds to fully non-conservative MT}
\tablenotetext{i}{MB prescription: ``IT03'' denotes that the \citet{IT2003} rate is used, while ``RVJ83'' corresponds to the \citet{RVJ1983} prescription}
\tablenotetext{j}{MB application criterion: ``$M_{2} < 1.5 \,\rm M_{\odot}$'' denotes that MB was applied to all binaries with donor mass less than $1.5 \,\rm M_{\odot}$, while ``conv. env.'' denotes that MB was applied only to binaries with a donor which has an outer convective envelope and an inner radiative core}
\end{deluxetable}

Fig.~\ref{MPI} shows the grid of initial BH and donor masses and orbital periods of the calculated evolutionary MT sequences within the bounds illustrated in Fig.~\ref{ParSp}. When MB is assumed to operate in \emph{all} stars with mass below $1.5\,M_{\odot}$, the parameter space of initial donor mass and orbital period at RLO that leads to evolutionary sequences  satisfying all observational constraints expands to higher donor masses and longer periods. We note that we were able to find MT sequences that satisfy the observational constraints using either MB law and application criterion, and assuming either conservative or non-conservative MT. Unfortunately this does not allow us to favor one MB or MT prescription over the other and thus we cannot put any limits on the theoretical models.

\subsection{Kinematic Evolution}

\begin{figure}[ht]
   \plotone{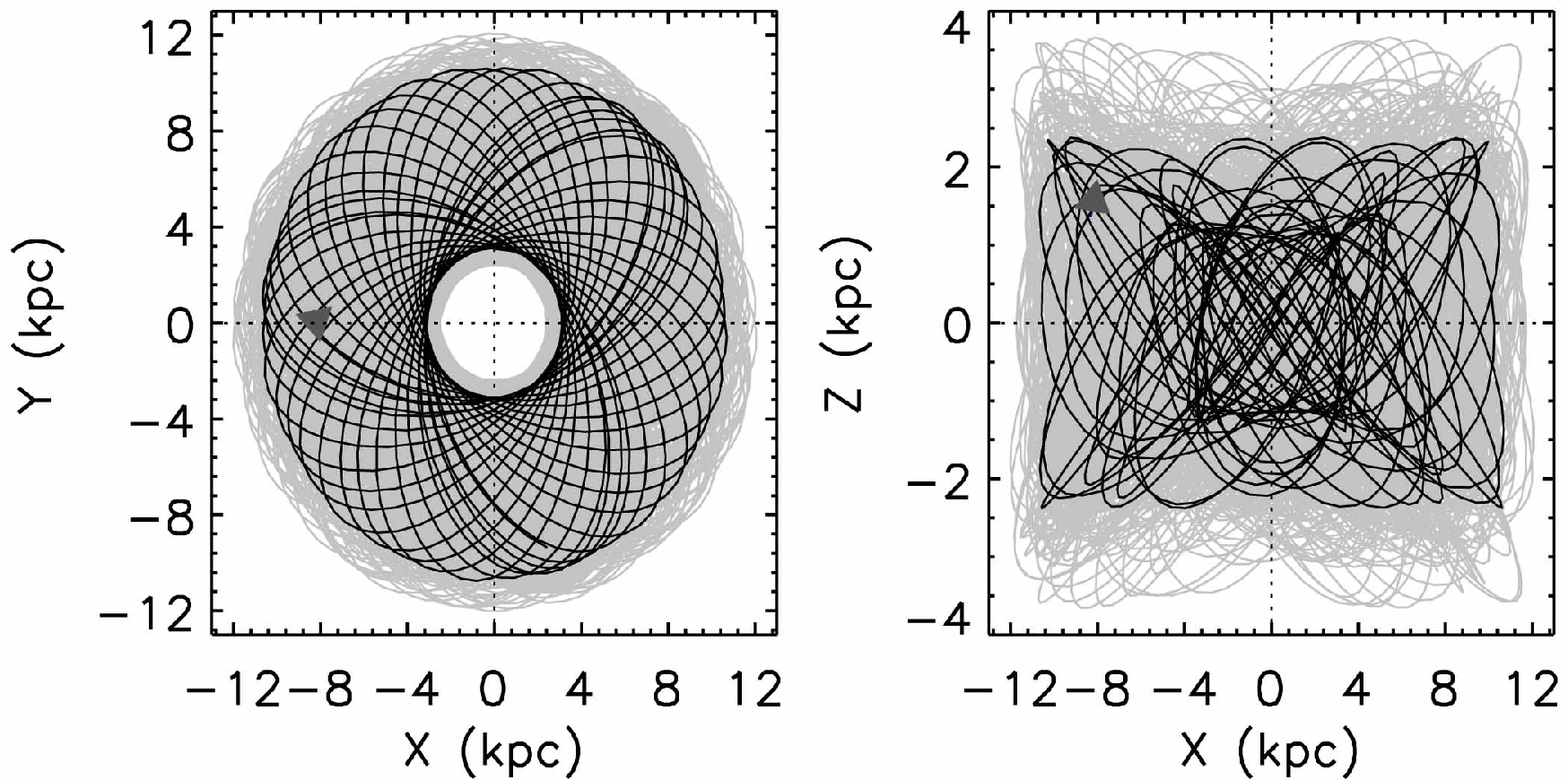}
   \plotone{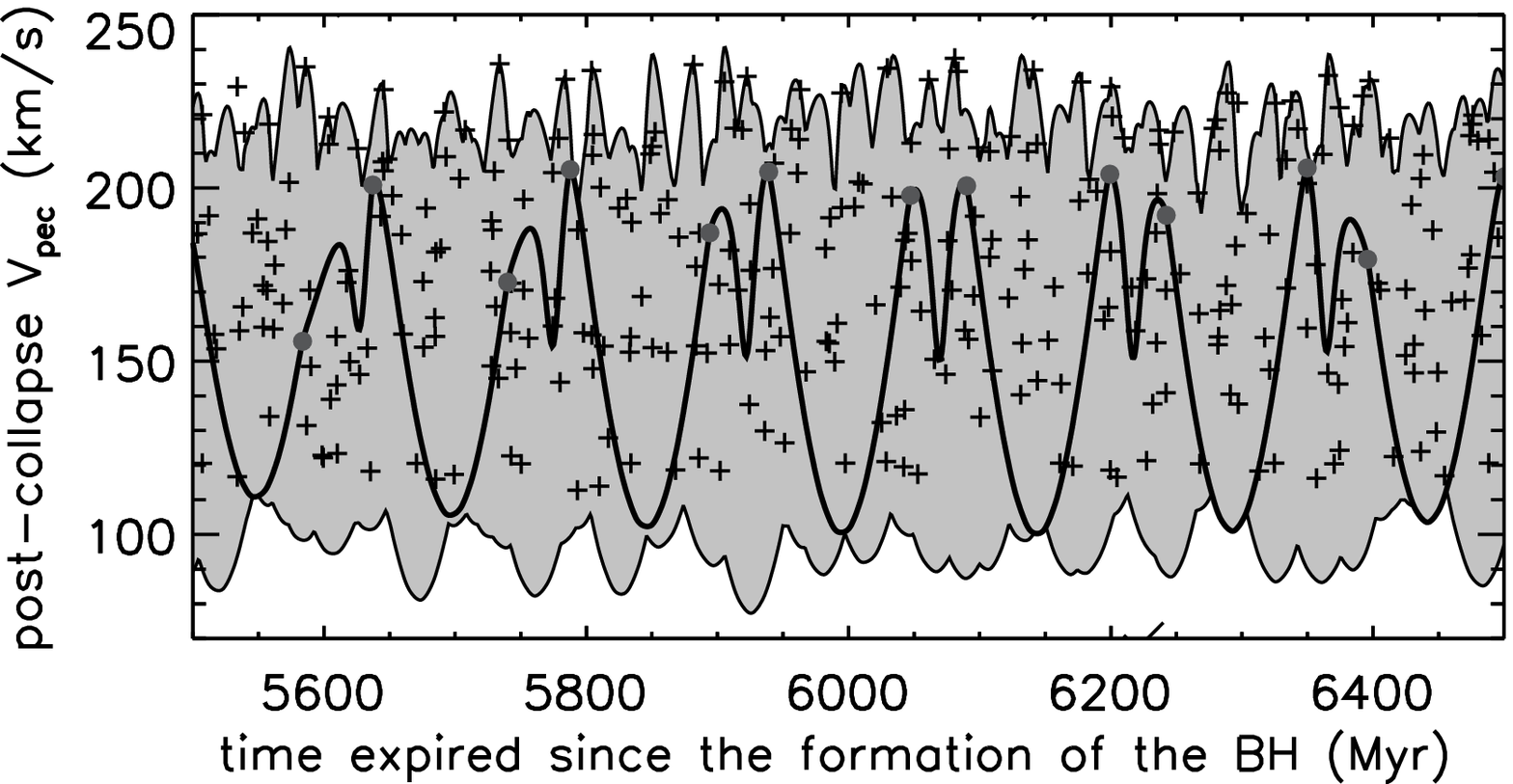}
   \caption{\emph{Upper panels:} Galactic orbit of XTE\,J1118+480 in the XY- and XZ-planes. The black line indicates the orbit for the mean values of the present-day distance and velocity components. The grey lines indicate the uncertainties in the orbit due to errors in the distance and the measured velocity components. The gray triangle indicates the present-day position of XTE\,J1118+480 . \emph{Lower panel:} Post-SN {\em peculiar} velocity of XTE\,J1118+480 as a function of the time expired since the formation of the BH. The black solid line represents the post-SN peculiar velocity for the mean values of the binary's current distance and velocity components, while the light grey area indicates the uncertainties resulting from the error bars in the these values. The dark gray filled circles indicate crossings of the orbit with the galactic plane, when we adopt the mean values of the distance and velocity, while the crosses take into account the associated uncertainties.}
\label{Motion}
\end{figure}

To model the Galaxy, we adopt the Galactic potential of \citet{CI1987} with updated model parameters of \citet{KG1989}. For each successful MT sequence, the equations governing the system's motion in the Galaxy are integrated backward in time up to the time corresponding to the current age of the donor star, as given by the evolutionary sequence. Our kinematic study shows that the system follows a quasi-periodic motion, crossing the Galactic plane approximately every 85\,Myr. The motion is bound in the Galactic potential covering values of R between 3\,kpc and 12\,kpc while oscillating in the Z direction with a maximum amplitude of 3.5\,kpc, where R is the radial distance from the Galactic center and Z is the vertical distance from the Galactic plane. The peculiar velocity of the system during the motion spans a wide range of values from 80\,$\rm{km\, s^{-1}}$ to 240\,$\rm{km\, s^{-1}}$ (see Fig.~\ref{Motion}). In all panels of Fig.~\ref{Motion}, black colors represent the orbit obtained for the mean values of the distance and velocity components, while grey indicates the deviations from this orbit obtained from considering all possible combinations of the extreme values of the distance and velocity component measurements (i.e. the endpoints of the error bars). Assuming that the system was born in the Galactic disk possible birthplaces are crossings of the orbit with the Galactic plane. We therefore identify the possible post-SN peculiar velocities by considering plane crossings that occur at a time in the past equal to the current age of the system (provided by the successful MT sequences) plus or minus 50\,Myr. Taking also into account the error bars in the measured current position and velocity of the system, allows us to derive lower and upper limits for the post-SN peculiar velocity (peculiar velocity of the system at possible BH birthplaces) for each successful MT sequence.

As mentioned above, we derive an age for the system from each successful MT sequence and use it to identify possible birth sites and set limits on the post-SN peculiar velocity $V_{\rm{pec,postSN}}$. Because the initial donor mass and orbital period at RLO of the successful MT sequences span a wide range ($M_2 = 1.0-1.6 \,\rm M_{\odot}$ and $P_{orb} = 0.5-0.8\, \rm days$), the ages of the system also span a wide range from 1.5 to 7\,Gyr. However, since the system follows a well-behaved quasi-periodic orbit that does not change characteristics over time, the post-SN peculiar velocity calculated for each MT sequence turns out to be insensitive to the age of the system. Combining all successful MT sequences, we find: $110\,\rm{km\, s^{-1}}<V_{\rm{pec,postSN}}<240\,\rm{km\, s^{-1}}$.

\subsection{Orbital Evolution Due to Tides and General Relativity}

After the formation of the BH and before the onset of the MT phase, the orbital parameters of the binary are subjected to secular changes due to the tidal torque exerted by the BH on the MS star
and due to the loss of orbital angular momentum via gravitational radiation and magnetic braking. Since the tidal interactions depend on both the orbital and rotational properties of the MS star, the star's rotational angular velocity right after the SN explosion that formed the BH enters the problem as an additional unknown quantity. Here we assume that the pre-SN orbit is circular, that the rotational angular velocity of the MS star is unaffected by the SN explosion, and that just before the explosion the star's rotation was synchronized with the orbital motion. The post-SN rotational angular velocity is then given by $\Omega_{\rm postSN} = 2\,\pi/P_{\rm orb,preSN}$, where $P_{\rm orb,preSN}$ is the pre-SN orbital period.  The system of equations governing the post-SN tidal evolution of the semi-major axis $A$, the orbital eccentricity $e$, and the MS star's rotational angular velocity $\Omega$ is given by \citet{Hut1981}: 

 \begin{eqnarray}
\lefteqn{\left( {{dA} \over {dt}} \right)_{\rm tides} = 
   -6\, {k_2 \over T}\, 
   {M_{\rm BH} \over M_2}\, {{M_{\rm BH}+M_2} \over M_2} 
   \left( {R_2 \over A} \right)^8}   \nonumber \\
 & & \times {A \over {\left( 1 - e^2 \right)^{15/2}}}
   \left[ f_1 \left( e^2 \right) - \left( 1 - e^2 
   \right)^{3/2} f_2 \left( e^2 \right) {\Omega \over n}
   \right],
\end{eqnarray}
\begin{eqnarray}
\lefteqn{\left( {{de} \over {dt}} \right)_{\rm tides} = 
   -27\, {k_2 \over T}\, 
   {M_{\rm BH} \over M_2}\, {{M_{\rm BH}+M_2} \over M_2} 
   \left( {R_2 \over A} \right)^8}   \nonumber \\
 & & \times {e \over {\left( 1 - e^2 \right)^{13/2}}}
   \left[ f_3 \left( e^2 \right) - {{11} \over {18}} 
   \left( 1 - e^2 \right)^{3/2} f_4 \left( e^2 \right) 
   {\Omega \over n} \right],  \label{dedt}
\end{eqnarray}
\begin{eqnarray}
\lefteqn{\left( {{d\Omega} \over {dt}} \right)_{\rm tides} = 
   3\, {k_2 \over T} 
   \left( {M_{\rm BH} \over M_2} \right)^2
   {{M_2\, R_2^2} \over I_2}
   \left( {R_2 \over A} \right)^6}  \nonumber \\
 & & \times {n \over {\left( 1 - e^2 \right)^6}}
   \left[ f_2 \left( e^2 \right) -  
   \left( 1 - e^2 \right)^{3/2} f_5 \left( e^2 \right) 
   {\Omega \over n} \right],
\end{eqnarray}
where $M_{\rm BH}$ is the mass of the BH, $M_2$, $R_2$, $k_2$, and $I_2$ are the mass, radius, apsidal-motion constant, and moment of inertia of the companion star, $n=2\,\pi/P_{\rm orb}$ is the mean orbital angular velocity, and $T$ is a characteristic time scale for the orbital evolution due to tides. The functions $f_i(e^2)$, $i=1,2,\ldots,5$, are defined by Eq.~(11) in  \citet{Hut1981}.

For stars with convective envelopes ($M_2 \lesssim 1.3\, \rm M_{\odot}$), the factor $ k_2/T $ can be approximated as
\begin{equation}
\left( \frac{k_2}{T} \right)_{\rm conv} = \frac{2}{21}f_{\rm cal}\frac{f_{\rm conv}}{\tau_{\rm conv}}\frac{M_{\rm env}}{M_2} \, \rm yr,	
\label{kTconv}
\end{equation}	
where
\begin{equation}	
\tau_{\rm conv} = 0.4311\left[ \frac{\frac{M_{\rm env}}{M_{\odot}}\frac{R_{\rm env}}{R_{\odot}}\left( \frac{R_2}{R_{\odot}}-\frac{1}{2}\frac{R_{\rm env}}{R_{\odot}}\right)  }{3\frac{L_2}{L_{\odot}}}  \right]^{1/3}\, \rm yr.
\label{Tconv}
\end{equation}	
Here $M_{\rm env}$ and $R_{\rm env}$ denote the mass and depth of the convective envelope, and $L_2$ denotes the bolometric luminosity of the donor star. The numerical factor $f_{\rm conv}$ is defined as
\begin{equation}
f_{\rm conv}=min\left[1,\left(\frac{P_{\rm tid}}{2\tau_{\rm conv}}\right)^2\right],
\end{equation}   
where the tidal pumping timescale is given by
\begin{equation}
\frac{1}{P_{\rm tid}}=\left|\frac{1}{P_{\rm orb}} - \frac{1}{P_{\rm spin}}\right|.
\end{equation}   
Here $P_{\rm spin}$ is the spin period of the star.

For stars with radiative envelopes ($M_2 \gtrsim 1.3\, \rm M_{\odot}$) the  factor $ k_2/T $ can be approximated as
\begin{eqnarray}
\left( {k_2 \over T} \right)_{\rm rad} & & = 
  1.9782 \times 10^4\, f_{\rm cal}
  \left( R_2 \over R_\odot\right)^2
  \left( R_\odot \over A \right) 
  \nonumber \\
 & & \times \left( M_2 \over M_\odot \right)
  \left( {{M_{\rm BH}+M_2} \over M_2} \right)^{5/6} 
  E_2 \,\, {\rm yr^{-1}},  \hspace{1.0cm}
  \label{kTrad}
\end{eqnarray}
where
\begin{equation}
E_2 = 1.592 \times 10^{-9} \left( M_2 \over M_\odot \right)^{2.84}
   \label{E2rad}
\end{equation}
\citep{Rasio1996,HTP2002,Belczynski2008}.

In the expressions for $k_2/T$, we incorporated a calibration factor $f_{\rm cal}$ to account for some of the still existing uncertainties in the strength of tidal dissipation. For stars with a convective envelope we adopt $f_{\rm cal}=50$, which yields tidal evolution timescales consistent with observationally inferred circularization periods of binaries in open clusters and measured orbital decay rates in high-mass X-ray binaries. For stars with radiative envelope, we set $f_{\rm cal}=1$ \citep[for more details see ][]{Belczynski2008}.

To follow the secular changes of the orbital parameters after the formation of the BH and before the onset of RLO, we integrate the equations governing the evolution of the orbit under the influence of tides, MB and general relativity forward in time. We note here that the form of Eq.~(\ref{dedt}) does not allow us to integrate backward in time, as a system with a circular orbit just before the onset of RLO would always remain circular when integrating backwards. For each successful MT sequence, we therefore consider pairs of post-SN orbital periods and eccentricities and integrate the set of orbital evolution equations forward in time until the secondary star fills its Roche-lobe. If the orbit has not yet circularized by that time, RLO will initially take place when the donor is at the periastron of the binary orbit. The integration is furthermore limited to a time interval less than or equal to the age of the donor star at the onset of RLO, as calculated from the MT sequence \footnote{In practice, we continue the integration for a time period 50\% longer than the age of the donor, as derived from the MT calculations. This way, we are able to account for a possible underestimate of the donor's age and thus for a miss-match of the time of RLO at periastron both before and after the time dictated by the successful MT sequences.}. 

By integrating the orbital evolution equations forward in time, we are able to map the post-SN orbital parameters to those at the onset of RLO. In particular, comparison of the orbital semi-major axis and eccentricity at the end of the orbital evolution calculation with the orbital parameters at the onset of RLO given by the successful MT sequences allows us to select those pairs of post-SN orbital period and eccentricity that yield the right initial conditions at the start of RLO to give rise to successful MT sequences. Since the binary stellar evolution code used to calculate the MT sequences assumes circular binary orbits, we match orbital configurations that are not circular at the end of the orbital evolution calculation by assuming instantaneous circularization while conserving the total angular momentum (orbital angular momentum plus spin angular momentum of the secondary star) of the system. However, the post-SN orbital evolution can lead to RLO at periastron at an age inconsistent with the age of the donor star at the onset of RLO as dictated by the successful MT sequences. So additionally, out of the pairs of initial orbital period and eccentricity that lead to orbital configurations consistent with the successful MT sequences at the onset of RLO, only those that lead to RLO at periastron at a time that is sufficiently close to  the age of the donor star required to make the sequence successful in the first place, are accepted.

\subsection{Supernova Dynamics and Progenitor Constraints}

To understand the core-collapse event leading to the formation of the BH, we are mainly interested in the constraints on the mass of the BH's helium star progenitor and the magnitude of the kick that may have been imparted to the BH at birth. At the time of the core collapse, the mass lost from the system and the possible natal kick imparted to the BH change the binary's orbital parameters. The pre- and post-SN component masses, orbital semi-major axis, and orbital eccentricity are related by the conservation laws of orbital energy and angular momentum, which depend on the magnitude and the direction of the kick velocity that may be imparted to the BH at birth \citep[for details see, e.g.,][]{Kalogera1996,FK1997}. The conservation laws provide two equations for five unknown parameters: the pre-SN orbital separation $A_{\rm preSN}$, the pre-SN mass of the BH's helium star progenitor $M_{\rm He}$, the magnitude of the kick velocity $V_k$, the polar angle $\theta$ between the kick velocity and the relative orbital velocity of the helium star just before the SN explosion, and the azimuthal angle $\phi$ defined so that $\phi=0$ represents a plane perpendicular to the line connecting the centers of mass of the binary components. The equations therefore provide a set of solutions rather than a unique solution for the pre-SN binary parameters and the possible BH natal kick velocity. We derive this set of solutions for each successful MT sequence, starting from the post-SN orbital parameters derived in the previous step. The solutions are constrained further by imposing that the post-SN center-of-mass velocity of the binary resulting from the mass loss and possible BH natal kick is within the range of post-SN peculiar velocities derived by tracing the motion in the Galaxy back in time. This procedure results in well defined regions of progenitor and kick properties that are consistent with all currently available observational constraints for XTE\,J1118+480. For more details, we refer to \citet{Willems2005}.  We note that, compared to \citet{Willems2005}, we here do impose an additional somewhat ad-hoc upper limit of $20\,\rm M_{\odot}$ on the mass of the BH's helium star progenitor, considering it as a physical limit on the maximum mass that a helium star can have (see \S\,6.1 for further discussion).

\begin{figure}[ht]
\plotone{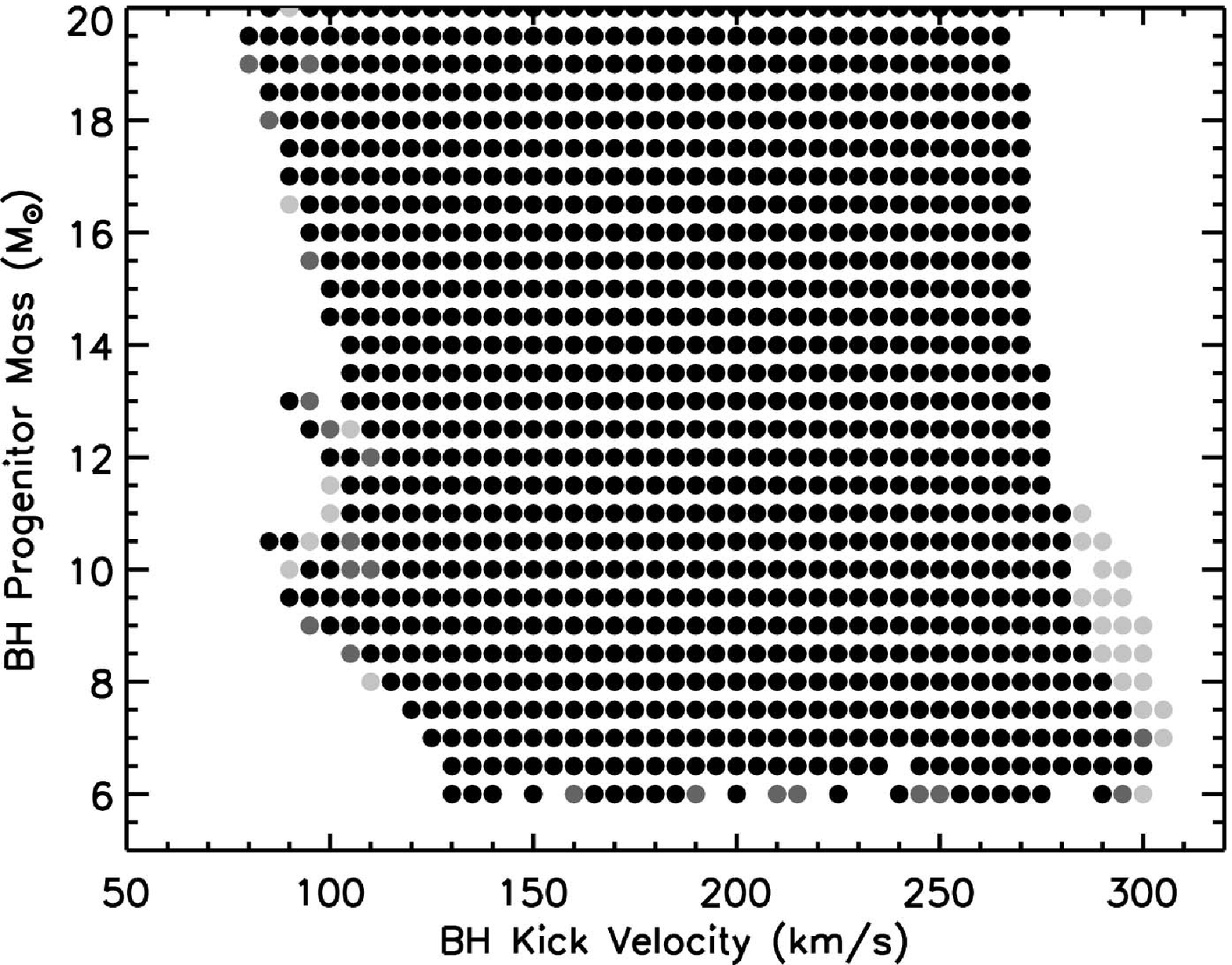}
\caption{Combined constraints on the mass $M_{\rm He}$ of the BH's helium star progenitor and the BH kick velocity magnitude $V_k$ from all successful MT sequences listed in Table \ref{sucMT}. The timing of RLO at periastron is either unconstrained (light gray dots), or constrained to be comparable to the age of the donor star (as provided by each successful MT sequence) within 250\,Myr (dark gray dots), or within 50\,Myr (black dots).}
\label{BHconstraints}
\end{figure}

In Fig.~\ref{BHconstraints}  we overlay the combined constraints on $M_{\rm He}$ and $V_k$ obtained from all successful MT sequences listed in Table \ref{sucMT} (i.e. for conservative as well as non-conservative MT sequences and for all adopted MB prescriptions). We find that the mass of the helium star is constrained to be larger than $6.5\,\rm{M_\odot}$ and that an asymetric natal kick is {\em required} to explain the formation of the system. The magnitude of the kick velocity is found to be in the range $80\,\rm{km\, s^{-1}} < V_{\rm{kick}}<310\,\rm{km\, s^{-1}}$.

The requirement that RLO, as predicted by the post-SN orbital evolution, occurs within a specified time window centered on the initial age of the donor star at the onset of RLO as given by the successful MT sequences does not significantly affect the constraints on $M_{\rm He}$ and $V_k$. This is illustrated in  Fig.~\ref{BHconstraints}, where the timing of RLO at periastron is either unconstrained (light gray dots) within the time of the integration\footnotemark[\value{footnote}], or constrained to be comparable to the age of the donor star (as provided by each successful MT sequence) within 250\,Myr (dark gray dots), or within 50\,Myr (black dots).

The insensitivity of our results to the imposed age constraints is in contrast to the findings of \citet{Willems2005} in their analysis of the evolutionary history of GRO\,J1655-40, where the requirement that RLO occurs at the right time had a major influence on the $M_{\rm He}$ and $V_k$ constraints. This difference comes from the fact that we are now dealing with lower mass donor stars ($\simeq 1.0-1.6\,M_{\odot}$). In this mass range, the secondary star may have a convective envelope in which case the tidal interactions are much stronger than when no such envelope is present. The orbit of systems containing a secondary star with a convective envelope therefore usually circularizes well before the onset of RLO. Synchronization of the secondary's rotation rate with the orbital motion generally takes place even sooner. For such synchronized and circularized binaries, our requirement that the total binary orbital angular momentum at the end of the orbital evolution calculations must match the total binary orbital angular momentum at the start of the successful MT sequences translates to matching, within the uncertainties of the calculation, the orbital separation at the end of the orbital evolution calculations with the orbital separation at the start of the successful MT sequences. In turn, this implies a matching of the Roche-lobe radius, which, at the onset RLO, is equal to the secondary star's radius. Since the radius of a MS star is a monotonically increasing function of time, our matching condition on the binary's total angular momentum in synchronized and circularized binaries is therefore equivalent to requiring RLO to occur at the right time. This equivalence does not hold for the non-synchronized, non-circular binaries which dominated the pre-MT progenitors of GRO\,J1655-40 in the analysis of  \citet{Willems2005}. Correspondingly, the differences between the grey and black dots in our Fig.\ref{BHconstraints} are caused by binaries with secondary stars more massive than $1.4\,M_{\odot}$ which have no convective envelope and are not able to achieve synchronization and circularization before the onset of RLO. Since these binaries cover only a small region of the parameter space, the differences between the parameter space represented by the grey and black dots are small.

An additional complication that enters the dynamical analysis of the core collapse comes from the fact that at the time of the BH formation ($\sim 5-10\,\rm Myr$ after the primary reached the ZAMS), the secondary star might still be a pre-MS star \citep{Ivanova2006}. In this case, the radius of the secondary is larger than the radius of a ZAMS star of the same mass. In order to take into account this scenario, we have to require that the post-SN orbital separation and eccentricity are such that the still pre-MS secondary star does not fill its Roche lobe at periastron, right after the BH formation. To examine the effect of this additional constraint due to the larger pre-MS star radius, we considered pre-MS models at $7\,\rm Myr$. We found that systems containing secondary stars with convective envelopes ($M_2\lesssim 1.3\,\rm M_{\odot}$), do not fill their Roche lobe right after the BH formation, as they usually have wide post-SN orbits which shrink later on due to the strong tidal interactions. Therefore, considering pre-MS secondary stars at the time of the BH formation does not affect our final constraints on the mass of the BH's helium star progenitor and the magnitude of the natal kick. On the contrary, systems containing secondary stars with radiative envelopes ($M_2\gtrsim 1.3\,\rm M_{\odot}$) have tighter post-SN orbits, as the weaker tides are not able to significantly alter the orbital characteristics. This leads to RLO right after the BH formation for many post-SN orbital configurations. In this case, the consideration of pre-MS secondary stars, eliminates all the post-NS orbits that lead to RLO at periastron, at a time close ($<250\,\rm Myr$ time offset) to that dictated by the successful MT sequences. In practice we do not know the exact time of BH formation, so that our choice of $7\,\rm Myr$ for the age of the pre-MS secondary stars right after BH formation is only an approximation. However, since the subset of systems containing secondary stars with radiative envelope covers only a small region of the $M_{\rm He}-V_k$ parameter space, the general shape of Fig.~\ref{BHconstraints} remains virtually unchanged.

\section{Three-Dimensional Core-Collapse Simulations}

The evolutionary calculations presented so far provide us with comprehensive and reliable constraints on the BH progenitor mass (in excess of 6.5\,M$_\odot$), the BH mass at the time of its formation ($6.0-10.0\rm \,M_\odot$) and the kick magnitude imparted to the BH at formation ($80-310\, {\rm km \, s^{-1}}$). In this section we examine whether current, state-of-the-art core-collapse simulations are consistent with the picture of BH formation as revealed by these constraints. We also investigate the nucleosynthetic yields from the simulations in the context of the observed abundances on the surface of the donor star. We stress, however, that we don't consider this study of core collapse simulations as a mean to derive strict constraints on the formation of the BH, since the computational cost of a systematic parameter study of the core collapse phase is prohibitive. We rather intend to demonstrate that we are able to unravel a self-consistent picture of the BH formation.   

\citet{Gonzalez2008}  presented a recent analysis of the abundance ratios for the elements Mg, Al, Si, Ca, Ti, and Ni in the companion star. They found that Ca and Ti are consistent with the average values of stars in this region.  Fe, Ni, and Si appear high, but are consistent with the average values at the 1$\sigma$ level.  Finally, Al appears high at the 2$\sigma$ level. Although an increase with respect to their stellar samples suggests that some enrichment did occur, these measurements provide a rather weak constraint on metal enrichment due to the supernova explosion that formed the BH in this system.

If indeed the system was formed at solar metallicity, the derived constraint on the BH mass at formation can be quite limiting. The reason is that at such metallicity mass loss from stellar winds limits the pre-collapse stellar mass to $\sim 10$M$_\odot$ (e.g. see results from \citet{Heg03,Young06}). Consequently the total mass that can be ejected at the explosion is also limited. Here we use the binary progenitors produced by \citet{Young08} (for the Cassioppeia A supernova remnant) \footnote{We note that other research groups \citep[e.g. ][]{MM2003} have predicted pre-collapse stellar masses of up to $\sim 17$M$_\odot$ for solar luminosity progenitors. However, we opted for the \citet{Young08} models, as they provide all the required information in a format that can be easily incorporated as initial conditions for our core collapse simulations.}. Specifically, for our fiducial core-collapse progenitor, we use a stellar model at an initial mass of 40\,M$_\odot$, which lost its hydrogen-rich envelope late in its evolution and ended its life as a 7.9\,M$_\odot$ helium-rich star. This is among the highest final masses at this metallicity, but clearly it is still not massive enough to explain the higher end of our initial BH mass range and allows for a maximum of $1.9\,\rm M_{\odot}$ of ejecta at the SN event. The requirement that a significant kick was imparted to the nascent BH, assuming it is associated with asymmetries in the explosion, would allow the ejection of heavy elements from deep inside the star while  still having a low total ejecta mass.

In the present simulations, we categorize the level of asymmetry in the collapse based on the opening angle of that asymmetry.  In our first set of simulations, we use a fairly narrow (30 degree) asymmetry imposed on a shock just above $7 \times 10^9$\,cm to correspond to typical hypernova asymmetries \citep{Nom05}.  Our second set of simulations is based on low-mode convection in the supernova \citep[e.g. ][]{Blo03, FY07,Fog07} with a 45 degree opening angle.  Using our 40\,M$_\odot$ progenitor star (7.9\,M$_\odot$ at collapse), we freely alter the level of the asymmetry.  For our hypernova simulations we include both 1 and 2-sided ``jets''.  In our low-mode convection explosions, we limit our models to 1-sided asymmetries.  Our suite of explosions, with their initial conditions, are described in Table~\ref{tab:3Dexp}. 

We simulate the explosions using the SNSPH smooth particle hydrodynamics code \citep{Fry06}.  We first use a 1-dimensional code \citep[see ][ for details]{Her94} to model the explosion out to 100\,s.  This structure is then mapped into our 3D SNSPH code.  In this 3-dimensional setup, we can implement the various asymmetries and follow the rest of the explosion.  Outside the jets or low-mode explosions, the expansion velocity is set to 1/10th the value of our 1-dimensional simulation.  In our jets or low-mode explosions, we vary the velocity to obtain a range of kicks, ejecta masses, etc. We do not model the neutron star itself, but place an absorptive boundary at 1/2 the initial inner particle radius to simulate the accretion of the infalling material onto the neutron star.  A  caveat on our yield results is that we do not include the effects  of the modified shock heating caused by varying these velocities. But fallback is far more important in determining the yields and  we believe this first-order approach is sufficient to draw conclusions  from the observed abundances.

Table~\ref{tab:3Dexp} summarizes the BH masses at formation and kick magnitudes for all of our simulations. It is interesting to note that a range of solutions are consistent with the initial-BH-mass and kick constraints derived from the evolutionary calculations, and hypernova energies are not required. 

Detailed nucleosynthetic yields could be used to place some additional constraints. However, given that the current overabundances are, except for Al, at the $1\sigma$ level, it is difficult to derive any reliable constraints. Also, the range of successful MT sequences suggests that the current BH donor has lost a significant amount of mass ($\gtrsim 0.8\,\rm M_{\odot}$) and hence it is unlikely that the current surface abundances still hold information about any enrichment at the time of the SN event. Nevertheless, we explore the implications of requiring some enrichment. First and foremost, even with asymmetric explosions, we require that the initial BH mass is at the low end of our mass range ($<7M_\odot$), if we are to get any enrichment from the supernova.    To compare the yields in Table~\ref{tab:3Dexp} at face value to the observations, let us first compare the elemental ejecta masses to the solar abundance counterparts: Y$_{Fe,\odot}$=0.018M$_\odot$, Y$_{Ca,\odot}$=$8.76\times 10^{-5}$M$_\odot$, Y$_{Ti,\odot}$=$3.86\times 10^{-6}$M$_\odot$, Y$_{Si,\odot}$=$10^{-3}$M$_\odot$, Y$_{Mg/Al,\odot}$=$10^{-3}$M$_\odot$.  These are the mass estimates of the abundances in the Sun.
    
One approach to compare the yields to the data is to assume that a fraction of the ejecta yield is mixed into the companion.  For example, in model 2Jet3.5n, $Y_{Ca}/Y_{Ca,\odot} \approx 2$ and $Y_{Mg/Al}/Y_{Mg/Al,\odot} \approx 7$.  If 10\% of the ejecta makes it into the companion, it would increase the Ca by 20\% and the Mg/Al by 70\%.  Probably the strongest constraint in the observations comes from the fact that Al in the companion is overabundant at the 2$\sigma$ level and Ca, if anything, is underabundant.  Comparing the  production of Mg/Al and Ca could well place a strong constraint.   The ratio of these abundances is also shown in Table~\ref{tab:3Dexp}.  A ratio value less than 1 indicates that the donor is enriched in Ca (above the solar value) more than it is enriched in Mg/Al.  Since the opposite is  true, we can rule out any model that has a Mg/Al ratio less than one.   Preferably, this ratio should be high, suggesting that models LM0.6, LM0.7 are among the best fits.  We can also rule out many of our jet models (where $Mg/Al<1$), but we cannot definitively  rule out model 2Jet3.5n.

\begin{figure}
\plotone{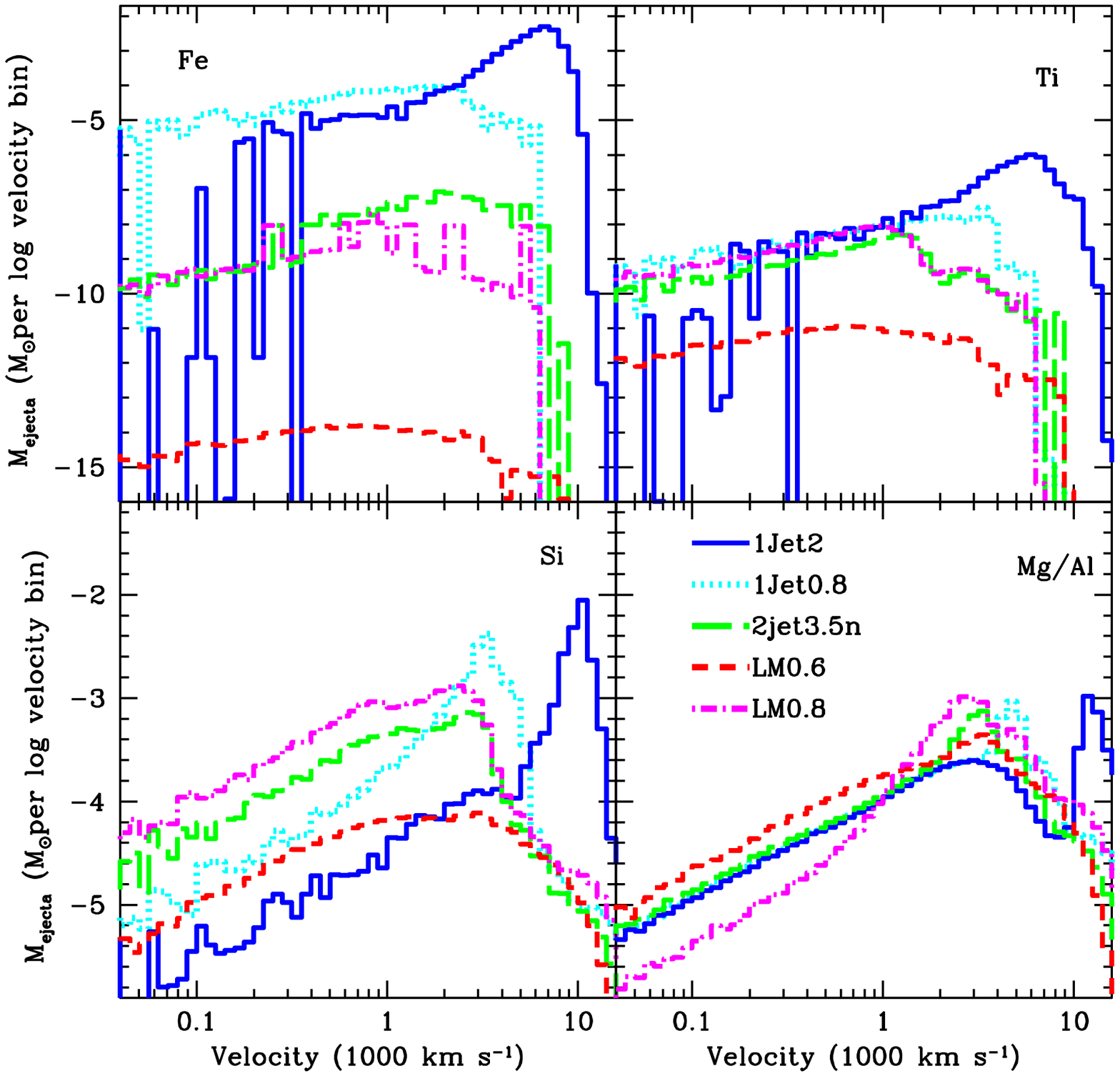}
\caption{Log of material abundances (in units of M$_\odot$ per logarithmically spaced velocity bin) versus velocity for iron (upper left panel), titanium (upper right panel),  silicon (lower left panel) and the total magnesium plus  aluminum (lower right panel).  We present the results  for 5 of our models:  1Jet2 (solid), 1Jet0.8 (dotted), 2Jet2 (long-dashed), LM0.6 (dashed), and LM0.8 (dot-dashed). The low velocity material is more likely to accrete onto the companion star.}
\label{fig:velabun}
\end{figure}

This analysis of the total yield leaves out a potentially important effect: only the slow moving ejecta material is accreted onto the companion. Figure~\ref{fig:velabun} shows the ejecta mass as a function of velocity.  Note that the LM0.6 model actually ejects more mass in Mg/Al below a velocity of 1000\,km\,s$^{-1}$ than any of the other models.  In contrast, it ejects much less Fe, Ti, and Si at these low velocities.  If anything, this effect argues more strongly that the low-mode convection models (in the LM0.4-LM0.7 range) are a better fit to the nucleosynthetic data.

In summary, both jet and low-mode convection simulations can be found that fit both the remnant mass and velocity measurements of this system.  But for either set of models, we require the mass of the black hole to be low ($<7,M_\odot$) to explain any nuclear enrichment in the companion.  We do not expect considerable enrichment in iron peak elements.  If we push the interpretation of the abundance data further to say that Al is enriched but not Ca, we conclude that the low-mode convection model explains all of the data better than a jet model, arguing against a hypernova progenitor for this system.  However, given the uncertainties in the abundance measurements, this last constraint is weak. We note here, that if we take the currently observed metallicity of the donor star at face value, and assume that the BH forming star had also super-solar metallicity, there is no need to consider chemical enrichment of the donor star in order to explain the currently observed spectrum. However, this makes the BH mass constraint even harder to satisfy, since there are no models in the literature that can produce massive BHs at super-solar metallicity. In this case the need for an asymmetry in the SN explosion and a large natal kick is even stronger.

\begin{deluxetable}{lcccccccccc}
\rotate
\tablewidth{583pt} 
\tablecaption{3-Dimensional Explosions \label{tab:3Dexp}}
\tablehead{ \colhead{Model} 
& \colhead{$f_{\rm velocity}$} 
& \colhead{$E_{\rm explosion}$}
& \colhead{initial $M_{\rm BH}$}
& \colhead{$V_{\rm kick}$}
& \colhead{$Y_{\rm MgAl}$}
& \colhead{$Y_{\rm Si}$}
& \colhead{$Y_{\rm Ca}$}
& \colhead{$Y_{\rm Ti}$}
& \colhead{$Y_{\rm Fe Peak}$} 
& \colhead{$R\tablenotemark{a}$} \\
\colhead{Name}
& \colhead{}
& \colhead{(10$^{51}$\,erg)} 
& \colhead{(M$_\odot$)}
& \colhead{(${\rm km s^{-1}}$)}
& \colhead{(M$_\odot$)} 
& \colhead{(M$_\odot$)} 
& \colhead{(M$_\odot$)} 
& \colhead{(M$_\odot$)} 
& \colhead{(M$_\odot$)} 
& \colhead{(M$_\odot$)} 
}

\startdata

1Jet2\tablenotemark{b} & 2.0 & 1.1 & 6.3 & 790 & 0.007 & 0.02 & 0.001 & $6\times10^{-6}$ & 0.02 & 0.6 \\
1Jet.8  & 0.8 & 0.30 & 6.4 & 300 & 0.007 & 0.02 & 0.0007 & $3\times10^{-7}$ & 0.001 & 0.9 \\
2Jet2  & 2.0,1.5 & 1.7 & 5.8 & 250 & 0.01 & 0.04 & 0.002 & $9\times10^{-6}$ & 0.04 & 0.4 \\
2Jet3.5n & 3.5,3.0 & 0.21 & 6.5 & 200 & 0.007 & 0.01 & 0.0002 & $5\times10^{-8}$ & $9\times10^{-7}$ & 3\\
LM0.2 & 0.2 & 0.083 & 6.8 & 20 & 0.004 & 0.0007 & $7\times10^{-5}$ & $3\times10^{-16}$ & $2\times10^{-20}$ & 5\\
LM0.4 & 0.4 & 0.13 & 6.6 & 80 & 0.005 & 0.0009 & $8\times10^{-5}$ & $1\times10^{-14}$ & $2\times10^{-20}$ & 5 \\
LM0.6 & 0.6 & 0.22 & 6.4 & 200 & 0.007 & 0.002 & $8\times10^{-5}$ & $2\times10^{-10}$ & $3\times10^{-13}$ & 8 \\
LM0.7 & 0.7 & 0.28 & 6.1 & 310 & 0.009 & 0.006 & $8\times10^{-5}$ & $1\times10^{-9}$ & $2\times10^{-12}$ & 10 \\
LM0.8 & 0.8 & 0.37 & 5.9 & 420 & 0.01 & 0.02 & $4\times10^{-4}$ & $1\times10^{-7}$ & $1\times10^{-7}$ & 2 \\

\enddata 
\tablenotetext{a}{$R=(Y_{\rm MgAl}/Y_{\rm MgAl,\odot})/(Y_{\rm Ca}/Y_{\rm Ca,\odot})$}
\tablenotetext{b}{We have 3 sets of models: one-sided narrow opening angle ($30^\circ$) termed 1Jetx, two-sided narrow opening angle ($30^\circ$) termed 2Jetx, and one-sided broad opening angle ($45^\circ$) termed LMx (Low-Mode explosions).  In all models, beyond the opening angle, we decrease the explosion velocity from our standard value by 0.1.  In the opening, we use the factor: $f_{\rm velocity}$.  Note that in 2Jet3.5n, we have a narrower opening angle ($15^\circ$).}

\end{deluxetable}

\section{DISCUSSION}

\subsection{Maximum Mass of the BH Progenitor}

In their analysis for GRO\,J1655-40 \citet{Willems2005} were able to place a strong constraint on the mass of the  helium star BH progenitor. The upper mass limit  came from the analysis of the orbital dynamics during the core-collapse event. In the case of XTE\,J1118+480, the same analysis does not give a clear upper mass limit for  the BH progenitor. This is because of the different configuration of XTE\,J1118+480 and mainly due to the operation of MB, which is able to shrink initially wide orbits on timescales shorter than the lifetime of the secondary star. However an upper mass limit  comes from the physics involved in the evolution of massive stars. In particular, \citet[][ see Fig.~6]{MM2003}, adopting a full spectrum of stellar wind prescriptions,  
found that the maximum helium star mass they could achieve, evolving stars at solar metallicity and with ZAMS mass up to $100\,\rm M_{\odot}$, was $\sim 15\,\rm M_{\odot}$ when they included moderate stellar rotation and  $\sim 17.5\,\rm M_{\odot}$ when they assumed no stellar rotation (both consistent with \citet{Heg03,Young06,Young08} models). Adopting either of these two upper limits for the maximum mass of the BH progenitor star, instead of our ad-hoc $\sim 20\,\rm M_{\odot}$ limit, does not affect our constraint on the magnitude of the SN kick, so that our results are robust with respect to the helium star upper mass limit.

\subsection{Globular cluster origin \label{GC}}

\citet{Mirabel2001} concluded a globular cluster origin of XTE\,J1118+480 to be more probable than a Galactic disk origin, as only an extraordinary kick from the SN explosion could have launched the black hole into the observed Galactic halo orbit from a birthplace in the disk of the Galaxy. However, a globular cluster origin of XTE\,J1118+480 generally implies a low metallicity donor star. Study of such MT sequences show  that the ones that successfully reproduce the observational properties usually have MT rates above or only marginally below the critical MT rate that separates transient from persistent behavior. In addition, the parameter space of donor mass and orbital period at the onset of RLO that leads to a system with the observed properties of XTE\,J1118+480 is even more constrained than what we showed in Fig.~\ref{MTseq} for solar metallicity. This is because we now have to account for an additional constraint on the current age of the donor star ($\gtrsim 10\,\rm Gyr$) so that it is consistent with the age of the halo stellar population. This in turns limits the mass of the donor star at the onset of the RLO to $\sim 0.9-1.0\,\rm M_{\odot}$.

The strongest argument against a globular cluster origin of XTE\,J1118+480 comes from chemical abundance analyses. \citet{Gonzalez2006} reported that the secondary star has a supersolar surface metallicity $[Fe/H]=0.2 \pm 0.2$, noting that a galactic halo origin of the system, although improbable, cannot be excluded, as the high surface metallicity might be due to pollution of the secondary star's surface during the SN event. Additionally, in our core collapse simulations, we had to require that the mass of the black hole is low ($<7\,M_\odot$) in order explain any nuclear enrichment in the companion star. Even in this case though, our successful MT sequences show that the donor star has lost at least $\sim 0.8\,\rm M_{\odot}$ of mass due to accretion onto the BH. So the outer layers of the star that could have been polluted from the SN ejecta are already removed and we shouldn't be able to observe the pollution effect now, unless the mixing in the donor star before the onset of MT was so strong that the heavy elements from the SN were transfered all the way to the inner layers of the star. Furthermore, in a subsequent work, \citet{Gonzalez2008} used a grid of SN explosion models and they found that metal-poor models associated with a formation scenario in the Galactic halo provide unacceptable fits to the observed abundances, while the best agreement between the model predictions and the observed abundances was obtained for metal-rich progenitor models, corresponding to a birth of the system in the thin Galactic disk. While this cannot formally exclude a globular origin, it renders such a scenario rather improbable, as there are only two Galactic globular cluster (Terzan 5 and Liller 2) with $[Fe/H] \gtrsim 0.0$ \citep{Harris1996}.

\subsection{CNO Processed Accreted Material}

\citet{Haswell2002}, using ultraviolet spectroscopy, found that the emission line strengths in XTE\,J1118+480 strongly suggest that the material accreting onto the BH has been significantly CNO processed. As a physical explanation of the observed spectrum they proposed  a scenario where the donor star exposed, due to the MT, its inner layers which have been mixed with CNO processed material from the central nuclear region. They concluded that the MT must have been initiated from a sufficiently massive ($\sim 1.5\,\rm M_{\odot}$) donor, so that the CNO process could change significantly the ratio of C to N to the very low value they observed. 

However, due to the lack of a precise determination of the current C to N ratio, we cannot set a clear lower limit on the initial donor mass based on the presence of CNO processed material. Our successful MT sequences suggest that the current mass of the donor star is in the range of $0.15$--$0.35\,\rm M_{\odot}$. We compared our findings with the online library of single star evolutionary sequences by \citet{Paxton2006}, which includes the chemical composition of stars in terms of fractional abundance as a function of the mass coordinate. We find that even for an $1.1\,\rm M_{\odot}$ secondary star, by the time of the onset of RLO, the ratio of C to N in the central $0.3\,\rm M_{\odot}$ is low enough ($log(C/N)\lesssim -1$) that it would be consistent with the observational findings of \citet{Haswell2002}. In contrast to the latter authors, we therefore did not set a lower limit of $\sim 1.5\,\rm M_{\odot}$ on the donor mass in our MT calculations but instead adopt a more conservative approach and study initial donor masses down to $1.0\,\rm M_{\odot}$.

\section{SUMMARY}

In this paper we constrained the progenitor properties and the formation of the BH in the XRB XTE\,J1118+480 assuming that the system originated in the Galactic disk and the donor had solar metallicity. We find that a high magnitude asymmetric natal kick is not only plausible but required for the formation of the system. The minimum kick ($80\,\rm km\,s^{-1}$) is significantly larger than that of the BH in GRO\,J1655-40 where the natal kick could be as low as a few tens of km/s or where the BH formation could even be consistent with a symmetric SN explosion in some cases. The BH in XTE\,J1118+480 is therefore the second one for which constraints favoring non-zero kick magnitudes are derived, but the first one for which we
can formally exclude a symmetric SN explosion.

Putting together the different pieces of our analysis we can uncover the complete picture of the evolutionary history of XTE\,J1118+480 back to the time of BH formation. The study of the MT sequences strongly constrains the properties of the binary at the start of its XRB phase. We find that at the onset of RLO the donor mass is 1.0-1.6\,$M_\odot$, the BH mass is 6-10\,$M_\odot$, and the orbital period is 0.5-0.8\,days. By calculating the secular evolution of the binary system after the formation of the BH until RLO starts at periastron, we mapped the post-SN orbital parameters to those at the onset of RLO. 

Finally, we examined whether current, state-of-the-art core-collapse simulations are consistent with the picture of BH formation as revealed by these constraints, investigating also the nucleosynthetic yields from the simulations in the context the observed abundances on the surface of the donor star. We found that both jet and low-mode convection models can be consistent with the remnant mass and velocity measurements of this system. For either set of models, we had to require that the mass of the black hole is low ($<7\,M_\odot$) in order explain any nuclear enrichment in the companion. Further interpreting the abundance data, we can say that Al is enriched but not Ca, concluding that the low-mode convection model explains all of the data better than a jet model, and arguing against a hypernova progenitor for this system. However, our successful MT sequences indicate that the current BH donor has lost a significant amount of mass ($\gtrsim 0.8\,\rm M_{\odot}$) and hence it is unlikely that the current surface abundances still hold information about any enrichment at the time of the SN event. 

The constraints on compact object progenitors and kicks derived from this analysis are of immense value for understanding compact object formation and exposing common threads and fundamental differences between black hole and neutron star formation. However, with only two systems analyzed so far we do not attempt to derive any global conclusions. In subsequent analyses, we intend to apply the procedure outlined above to a number of additional XRBs: Cyg X-1, LS\,5039, LSI\,$+61^\circ\,303$, Vela\,X-1, 4U1700-37, Sco\,X-1. By examining both NS and BH systems and both RLO and stellar wind driven MT, we hope to unravel the systematic dependencies between the masses of newly formed compact objects and their immediate pre-SN progenitors, the mass lost at core collapse, and the possible kick velocity imparted to the compact object at birth.

\acknowledgments 

We are indebted to Laura Blecha for sharing the code used to follow the motion of XTE\,J1118+480 in the Galactic potential. This work is supported by a Packard Fellowship in Science and Engineering grant and an NSF CAREER award to VK. This project was funded in part under the auspices of the U.S. Dept. of Energy, and supported by its contract W-7405-ENG-36 to Los Alamos National Laboratory, and by NASA grant SWIF03-0047.


\end{document}